\documentclass[twocolumn]{aastex62}

\usepackage{natbib}
\usepackage{url}
\usepackage{academicons}
\usepackage{xcolor}
\usepackage{hyperref}
\usepackage[mathscr]{euscript}
\usepackage{comment}
\usepackage{amsmath}
\usepackage{amsbsy}
\usepackage{bm}

\usepackage{lineno}
\usepackage[colorinlistoftodos,textwidth=15mm]{todonotes}
\DeclareUnicodeCharacter{2212}{\textendash} % necessary for arXiv

\ifxetex
  \usepackage{letltxmacro}
  \LetLtxMacro\SavedIncludeGraphics\includegraphics
  \def\includegraphics#1#{% #1 catches optional stuff (star/opt. arg.)
    \IncludeGraphicsAux{#1}%
  }%
  \newcommand*{\IncludeGraphicsAux}[2]{%
    \XeTeXLinkBox{%
      \SavedIncludeGraphics#1{#2}%
    }%
  }%
\fi

\definecolor{orcidlogocol}{HTML}{A6CE39}

\shorttitle{Software for Metallicity Calibration of RR Lyrae stars}
\shortauthors{Spalding et al.}
\begin{document}
\defcitealias{clementinietal1995}{C95}
\defcitealias{lambertetal1996}{L96}
\defcitealias{fernleyetal1997}{F97}
\defcitealias{solanoetal1997}{S97}
\defcitealias{wallerstein2010composition}{W10}
\defcitealias{liu2013abundances}{L13}
\defcitealias{nemecetal2013}{N13}
\defcitealias{pancino2015chemical}{P15}
\defcitealias{chadid2017spectroscopic}{C17}
\defcitealias{layden1994}{L94}
\defcitealias{kemper1982}{K82}
\defcitealias{sneden2017rrc}{S17}
\defcitealias{crestani2021use}{C21}

\title{\texttt{rrlfe}: Software for Generating and Applying Metallicity Calibrations for RR Lyrae Variable Stars Across a Wide Range of Phases and Temperatures}

\author[0000-0003-3819-0076]{Eckhart Spalding \href{https://orcid.org/0000-0003-3819-0076}{\includegraphics[scale=0.1]{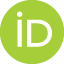}}}
\email{eckhart.spalding@sydney.edu.au}
\affil{Sydney Institute for Astronomy, School of Physics, The University of Sydney, Sydney, NSW 2006, Australia}
\affil{Department of Physics and Astronomy, University of Notre Dame, 225 Nieuwland Science Hall, Notre Dame, IN 46556, USA}

\author[0000-0002-4792-7722]{Ronald Wilhelm \href{https://orcid.org/0000-0003-3819-0076}{\includegraphics[scale=0.1]{images/orcid_64x64.png}}}
\affil{Department of Physics and Astronomy, University of Kentucky, Lexington, KY 40506, USA}

\author[0000-0002-3657-0705]{Nathan De Lee \href{https://orcid.org/0000-0002-3657-0705}{\includegraphics[scale=0.1]{images/orcid_64x64.png}}}
\affil{Department of Physics, Geology, and Engineering Technology, Northern Kentucky University, Highland Heights, KY 41099}
\affil{Department of Physics and Astronomy, Vanderbilt University, 301 Stevenson Center Ln., Nashville, TN 37235, USA}

\author[0000-0002-0726-424X]{Stacy Long
\href{https://orcid.org/0000-0002-0726-424X}{\includegraphics[scale=0.1]{images/orcid_64x64.png}}}
\affil{Department of Chemistry and Physics, Lincoln Memorial University, Harrogate, TN, 37752, USA}

\author[0000-0003-4573-6233]{Timothy C. Beers
\href{https://orcid.org/0000-0003-4573-6233}{\includegraphics[scale=0.1]{images/orcid_64x64.png}}}
\affil{Department of Physics and Astronomy and JINA Center for the Evolution of the Elements, University of Notre Dame, 225 Nieuwland Science Hall, Notre Dame, IN 46556, USA}

\author[0000-0003-4479-1265]{Vinicius M. Placco
\href{https://orcid.org/0000-0003-4479-1265}{\includegraphics[scale=0.1]{images/orcid_64x64.png}}}
\affil{NSF's NOIRLab, Tucson, AZ 85719, USA}

\author[0000-0003-0497-2651]{John Kielkopf \href{https://orcid.org/0000-0003-0497-2651}{\includegraphics[scale=0.1]{images/orcid_64x64.png}}}
\affil{Department of Physics and Astronomy, University of Louisville, Louisville, KY 40292, USA}

\author[0000-0001-5297-4518]{Young Sun Lee \href{https://orcid.org/0000-0001-5297-4518}{\includegraphics[scale=0.1]{images/orcid_64x64.png}}}
\affil{Department of Astronomy and Space Science, Chungnam National University, Daejeon 34134, Republic of Korea}
\affil{Department of Physics and Astronomy, University of Notre Dame, 225 Nieuwland Science Hall, Notre Dame, IN 46556, USA}

\author[0000-0002-3827-8417]{Joshua Pepper \href{https://orcid.org/0000-0002-3827-8417}{\includegraphics[scale=0.1]{images/orcid_64x64.png}}}
\affil{Department of Physics, Lehigh University, Bethlehem, PA 18015, USA}

\author[0000-0002-6307-992X]{Kenneth Carrell
\href{https://orcid.org/0000-0002-6307-992X}{\includegraphics[scale=0.1]{images/orcid_64x64.png}}}
\affil{Department of Physics and Geosciences, Angelo State University, San Angelo, TX 76909, USA}

\begin{abstract}
RR Lyrae stars play a central role in tracing phase-space structures within the Milky Way because they are easy to identify, are relatively luminous, and are found in large numbers in the Galactic bulge, disk, and halo. In this work, we present a new set of spectroscopic metallicity calibrations that use the equivalent widths of the {C}a II {K} and Balmer {H}$\gamma$ and {H}$\delta$ lines to calculate metallicity values from low-resolution spectra. This builds on an earlier calibration from Layden by extending the range of equivalent widths which map between Ca II K and the Balmer lines. We have developed the software \texttt{rrlfe} to apply this calibration to spectra in a consistent, reproducible, and extensible manner. This software is open-source and available to the community. The calibration can be updated with additional datasets in the future.
\end{abstract}


\section{Introduction}

RR Lyrae stars are old ($\gtrsim$10 Gyr), mostly iron-poor, helium-burning stars that exhibit regular luminosity variations as they pulsate. They are of central importance in tracing out phase-space structures in the Milky Way because they are numerous, relatively luminous ($+0.4\lesssim\langle M_{V}\rangle\lesssim+0.9$; \citealt{smith1995}), easy to identify based on their light-curve shapes, and are detectable deep into the Galaxy's baryonic halo, to $\sim$100 kpc from the Galactic centre and beyond.

The weak tidal gradients and shallow gravitational potential of the Galactic halo allow phase-space structures to persist for a few billion years, providing a fossilized record of the Milky Way's assembly history \citep{wettereretal1996,amrose2001calculation,helmi2008,iorio2021chemo}. Halo studies were used to investigate this history beginning with \citet{eggenetal1962}, who concluded that the Milky Way had undergone a rapid collapse that left behind a relatively homogeneous halo. This paradigm was challenged by \citet{searleetal1978}, who found that the heterogeneous metallicities and horizontal-branch morphologies of globular clusters suggested that the history of the halo involved a more protracted hierarchical accretion. \citet{carolloetal2007}, \citet{carolloetal2010}, \citet{beersetal2012}, and others since, have demonstrated the existence of at least two populations of halo stars, which they identified as the inner-halo population and the outer-halo population, with distinct spatial density profiles, stellar orbits, and stellar metallicities.  They took this as evidence that the 
individual halo components likely were formed in fundamentally different ways, e.g., 
through successive dissipational (inner) and dissipationless (outer) mergers and tidal disruption of proto-Galactic clumps. 
The emerging picture of the Milky Way's outskirts ($\gtrsim$15 -- 20 kpc; \citealt{deason2011milky,navarro2021rr}) has indeed confirmed a ``lumpy'' appearance, with numerous halo over-densities, including some two dozen known stellar streams \citep{mateu2017fourteen,bonaca2018information,shipp2018stellar,prudil2021milky} left over from the tidal disaggregation and accretion of dwarf galaxies. 

In addition to being interesting in its own right as a record of the Milky Way's assembly history, halo substructure also provides a probe of $\Lambda$CDM cosmology at galactic ($\sim$kpc) scales \citep{helmi2008, johnston2008tracing,cooper2010galactic}, where the physics of $\Lambda$CDM become complicated by stellar feedback in the form of heating, ionization, and metal enrichment \citep{springeletal2006}. The physics to be gleaned may contribute to the resolution of longstanding conundrums such as the ``missing satellites problem,'' whereby the dozens of known dwarf galaxies surrounding the Milky Way are much fewer than the hundreds expected from $\Lambda$CDM simulations (e.g., \citealt{bullock2017small}). Further progress on questions relevant to halo substructure and the Galactic potential will require tight constraints on the phase-space, age, and chemical information content of the diffuse halo and multiple stellar streams (e.g., \citealt{sanderson2017modeling,ramos2020full}). 

RR Lyrae stars are located at the intersection of the instability strip and the Horizontal Branch on the HR diagram. As lower-mass counterparts to Cepheid variables, RR Lyrae stars have a long history of use as standard candles for mapping the structure, kinematics, and formation history of the Milky Way (e.g., \citealt{vivas2008spectroscopy,braga2015distance,pietrukowicz2015deciphering,minniti2016discovery,dong2017near,mateu2017fourteen}). They include the subtype ab, which pulsates in the fundamental mode; subtype c, which pulsates in the first overtone mode; and subtype d, which pulsates in both modes. The ab and c subtypes (RRab and RRc stars) occupy distinct regions of period-amplitude space, whereby RRab stars are identifiable by qualitatively ``shark-tooth'' light curves, pronounced atmospheric shock waves, and cooler effective temperatures. RRc stars have a more sinusoidal light curves, and hotter effective temperatures \citep{catelan2015pulsating}.

Since the advent of large-scale variability surveys in the late 1990s and early 2000s, RR Lyrae stars have been unearthed by the thousands (e.g., \citealt{kinemuchi2006analysis,palaversa2013exploring,zinnetal2014,hernitschek2016finding,gran2016mapping,soszynski2016ogle,cohen2017outer,drake2017catalina,minniti2017characterization,jayasinghe2018asas,stringer2019identification,stringer2021identifying}). Their sheer numbers have enabled the tracing of stellar density profiles and structure out to beyond 100 kpc, deep into the halo \citep{fiorentino2014weak,hendel2017smhash,hernitschek2017geometry,sanderson2017new,sesar2017100,cohen2017outer,belokurov2017unmixing,hernitschek2018profile,martvaz2019}, and have mapped out dwarf galaxies in various stages of cannibalization \citep{sesar2014stacking,baker2015charting,martinez2015variable,molnar2015pushing,medina2017serendipitous,ferguson2019three,muraveva2020fresh}.

The dominant drivers of RR Lyrae pulsations are the effect of the increasing opacity as a function of temperature at certain ionization zones, particularly that of the second ionization layer of He (the $\kappa$-mechanism)  and the increased flow of heat into those same zones (the $\gamma$-mechanism). Within the instability strip, RR Lyrae stars rove back and forth on the HR diagram over a temperature range of T$_{\rm eff}$ $\sim$ 6000\,K to 7250\,K \citep{catelan2015pulsating}. 

In addition to their utility as standard candles, RR Lyrae stars have served as markers of metallicity. Given the long lifetimes of RR Lyrae stars, their iron abundances constrain the timescales, relative chronologies, and variegated parenthoods of stellar populations, in ways that are complementary to 6D phase-space information. 

Within the temperature range of RR Lyrae stars, the Balmer lines become stronger with higher temperatures, as more electrons are available in the first excited states for the absorption of photons, and temperatures are not hot enough for the effects of ionization to dominate. Thus, the Balmer-line equivalent widths (EWs) are strongly correlated with effective temperature. Over the same temperature range, the Ca II lines weaken with increasing temperature as they ionize, but also possess some correlation with iron abundances \citep{gray2009stellar,fernandez2015deep}. The inverse behavior of the changing Balmer and Ca II lines, and the partial correlation of Ca II K with iron, offers the possibility of calculating RR Lyrae iron abundances with multi-epoch spectroscopy. 

\citet{preston1959} was the first to note that the metallicity of RR Lyrae stars appeared to correlate with differences in spectral type, as determined from the Ca II K and hydrogen Balmer lines at minimum light. 
Later updates to the method of Preston (e.g., \citealt{butler1975,freeman1975chemical}) led to the calibration of \citet{layden1994}, henceforth \citetalias{layden1994}, which was based on the pseudo-equivalent widths of the Ca II K and Balmer H$\delta$ and H$\gamma$ lines. However, since the \citetalias{layden1994} calibrations were based on spectra at low-temperature phases (and thus narrow Balmer-line EWs), the calibration breaks down at hotter temperatures. Furthermore, the \citetalias{layden1994} calibrations with interstellar calcium correction were based on a model based on distance from the Galactic plane, symmetric in Galactocentric azimuth \citep{beers1990estimation}. Since then, it has become increasingly clear that interstellar Ca II absorption varies significantly along different lines-of-sight (e.g., \citealt{welty1996high}). Thus, the time is ripe for a new calibration which both maximizes the range of usable light-curve phases, takes a new look at interstellar Ca correction, and will be applicable to large spectroscopic datasets. In addition, none of the calibration stars used for setting iso-metallicities was more metal rich than [Fe/H] = $-0.34$, as calculated in \citetalias{layden1994}. A calibration that includes more metal-rich stars would be applicable to many additional stars in the halo and disk systems. Given the range in effective temperatures of these stars, a new relation between EWs can also be made to allow for the non-linear behavior of the lines.

Fortuitously, RR Lyrae stars are plentiful in the Sloan Digital Sky Survey (SDSS) dataset \citep{sesaretal2007}, which has yielded millions of low-resolution spectra since the first data release in 2003 \citep{abazajian2003first}. Though not designed as a variable star survey \citep{castander1998}, a subset of perhaps thousands of SDSS spectra represent randomly phased, multi-epoch spectra of RR Lyrae stars. Given the survey sensitivity, RR Lyrae stars are detectable in SDSS data out to distances $\gtrsim$100 kpc \citep{newberg2003, ivezicetal2005, sesar2011}, and have been used since the early days of the SDSS to trace out the physical and radial velocity structure of halo over-densities (e.g., \citealt{ivezicetal2000, sesaretal2007, belletal2008, watkinsetal2009, sesaretal2010, simionetal2014,casetti2015kinematically}).

The SEGUE Stellar Parameter Pipeline (SSPP) was designed to calculate, among other things, stellar metallicities based on SDSS spectra \citep{lee2008segue,lee2008segue2,prieto2008segue,smolinski2011segue}. However, the SSPP assumes LTE conditions, which are not always a fair approximation in pulsating variable stars. 

Photometric RR Lyrae metallicity calibrations also exist in the form of metallicity-period-$\phi_{31}$ relations, where $\phi_{31}$ is the Fourier phase parameter relation \citep{jurcsik1996determination,sandage2004,ngeow2016palomar,hajdu2018data,lietal2023}. Our spectroscopic calibration will help provide a basis for photometric metallicity calibrations, which will undergo renewed importance in the era of the photometric Legacy Survey of Space and Time (LSST) at the Vera C. Rubin Observatory. After routine science observations begin in the near future, LSST will detect millions of RR Lyrae stars out to hundreds of kpc to even a few Mpc, to the farthest outskirts of the Galactic halo and even beyond to other members of the Local Group \citep{oluseyi2012simulated,baker2015charting,sanderson2017new}. It will be impractical to obtain follow-up spectroscopy on such a large number of stars, especially since LSST goes down to $r\sim 24.5$ for single visits and for some observations down to $r\sim 26$ \citep{abell2009lsst}. Thus, a recalibration of photometric techniques will be important.

Here we present an updated metallicity calibration for RR Lyrae stars as generated from a basis set of synthetic spectra and empirical RRab spectra, and provide the software for applying this calibration to other stars and for revising this calibration in the future. To generate the calibration, our software normalizes the synthetic spectra, measures the EWs of the Ca II K, H$\gamma$, and H$\delta$ lines, and generates a functional fit. A final correction is applied, based on the comparison of retrieved values of [Fe/H] with a sample of RR Lyrae stars with a wide range of metallicities, as determined from high-resolution spectroscopy from previous studies. 

To obtain [Fe/H] estimates  with our pipeline, we collected multi-epoch spectra for stars at $R\sim2,000$ to mimic those in SDSS, as 
spectra for our program stars do not exist in SDSS as of DR14. We phase them using photometry from TESS, KELT, and other sources, so as to determine where in the phase the effects of shocks may cause the calibration to break down.  The resulting fits can be applied to single-epoch RR Lyrae spectra from SDSS or other surveys. \par

The organization of our paper is as follows. We describe our methods in Sec.\  \ref{sec:secMethods}, and the reduction of observations in Sec.\  \ref{sec:secAnalysis}. Results are described in Sec.\  \ref{sec:secResults}, which includes a test of our calibration on empirical spectra and the 
well-studied SDSS Stripe 82 in Sec.\  \ref{subsec:stripe82}. We provide a discussion of the implications in Sec.\  \ref{sec:secDiscussion}, and summarize in Sec.\  \ref{sec:secSummary}.

\section{Methods}
\label{sec:secMethods}

\subsection{Synthetic Spectra}
\label{subsec:synth_gen}

We computed a grid of synthetic spectra which covers the range in expected physical parameters for RR Lyrae stars.  The grid spans $5750\,K \le \textrm{T}_{\textrm{eff}} \le 7750$\,K, in increments of 250\,K, $2.0 \le$ log g $\le 3.0$, in increments of 0.5 dex, and $-2.5 \le [\textrm{Fe/H}] \le 0.0$, in increments of 0.5 dex.  The synthetic spectra were generated using \citet{castelli_kur_2003} Atlas9-ODF, LTE model atmospheres with microturbulence of 2.0 km s$^{-1}$, and the spectral synthesis code \texttt{SPECTRUM} v2.76 \citep{gray1994calibration}.  The synthetic spectra were generated at high resolution and Gaussian smoothed to a resolution of 3.2\,{\AA} with a dispersion of 1.4\,{\AA} per pixel.  Since we are using the Ca II K line EW as a proxy for Fe abundance, we scaled the calcium abundance at low metallicity using the relation from Eqn.~2 in \citet{prieto2006spectroscopic}.  The spectra span a range of 3900\,{\AA} $\le \lambda \le 5000$\,{\AA}, which included the lines of Ca II K and H, and the Balmer lines of H$\beta$, H$\gamma$, H$\delta$, and H$\epsilon$. These spectra were then processed by our custom pipeline \texttt{rrlfe}, which we describe in Sec. \ref{subsubsec:syntheticSpectra}.

\subsection{Literature Metallicities}
\label{subsec:lit_metal}

Estimates of [Fe/H] based on synthetic spectra must be compared with high-resolution studies of actual stars. We selected a set of 19 program stars for comparison and observation, including both RRab and RRc stars, most of which have metallicities calculated from high-resolution studies. These literature sources of [Fe/H] are listed in the appendix in Table \ref{table:sources}. We derived a net [Fe/H] value from multiple appearances of each star in the literature, following the general method of \citet{chadid2017spectroscopic}, which we summarize as follows.

\begin{figure}
\centering
\includegraphics[trim={0.8cm 0cm 1.3cm 0.5cm}, clip=True, width=1.0\linewidth]{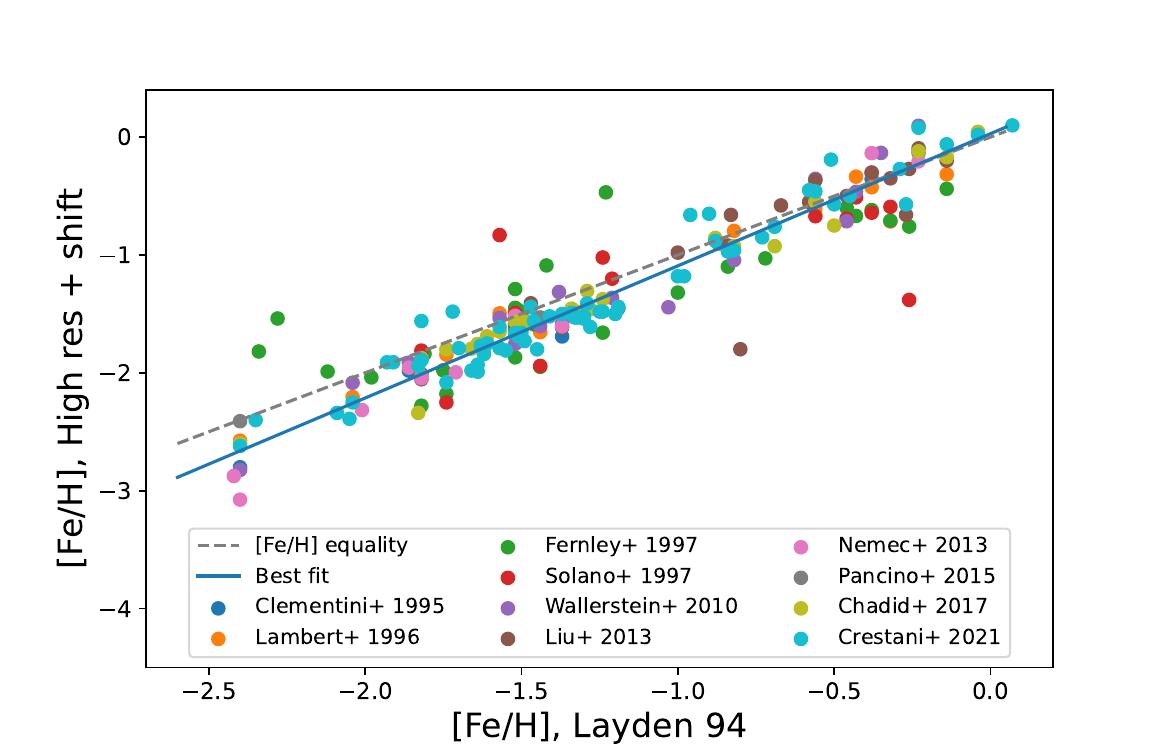}
\caption{[Fe/H] values from literature sources based on high-resolution spectroscopy of RRab stars, as mapped onto \citetalias{layden1994}. Offsets have been applied to each high-resolution literature source to minimize systematics. (See Fig.\ 7 in \citealt{chadid2017spectroscopic}.)}
\label{fig:high_res_ab_basis}
\end{figure}

Metallicity values from a given high-resolution study are plotted against those of \citetalias{layden1994}, for stars shared between both works. \citetalias{layden1994} included only RRab stars and was not a high-resolution study, but involved over 300 RR Lyrae stars, increasing the probability of sample overlap. The offset between a line of best fit at an arbitrarily-chosen abscissa value of ${\rm [Fe/H]}_{L94}=-1.25$, and the ordinate of the linear fit to the metallicities of 26 program stars of \citet{chadid2017spectroscopic}, is added in to shift all the high-resolution [Fe/H] values from a given study. After this is done for each of the high-resolution studies, this acts to remove systematic differences between them and maps them onto a common scale, while preserving the intrinsic scatter in metallicity among stars within each study as compared to \citetalias{layden1994}. Fig.\ \ref{fig:high_res_ab_basis} shows these shifted literature values against \citetalias{layden1994}.

We considered using a second-order polynomial or a piecewise-linear function to fit the metallicities from high spectral-resolution studies as a function of the \citetalias{layden1994} values. However, we found that the scatter at the metal-poor end did not provide better fits, at the price of an additional degree of freedom. This left us with a net literature metallicity value $\textrm{[Fe/H]}_{lit}$ based on the linear fit $\textrm{[Fe/H]}_{lit} = A\, {\rm [Fe/H]}_{L94} + B$, with $A=1.040\pm 0.026$ and $B=- 0.060 \pm 0.035$. Metallicities mapped in this way, along with their literature values, are tabulated in Table \ref{table:prog_star_fehs}. It should be noted that this mapping is generated using only RRab stars. To our knowledge, the largest single compendium of RRc metallicities is \citet{kemper1982}, which used the $\Delta S$ method on image tube scans of RRc stars observed from Lick Observatory.  The number of RRc stars which have published high-resolution spectroscopy is small, and only four such stars overlap with \citet{kemper1982}: U Com, T Sex, DH Peg, and YZ Cap. We therefore refrain from making another metallicity mapping for RRc stars alone, and apply any resulting calibration to them as-is.

\subsection{Photometry}

The very pulsation that causes effective temperature changes in RRab sends shocks through the photosphere, which can perturb the line profiles \citep{chadid1996turbulence,fokin1999shock,chadidetal2008,preston2011,chadid2017spectroscopic}. To put a constraint on this region of the phase in our program stars, as well as to make a general prediction of the phase for planning purposes, we initially used data from the website of the American Association of Variable Star Observers (AAVSO) \citep{aavso_ack}. To minimize the effect of drift in predicted phase we obtained follow-up observations from December 2012 to January 2014, from three locations: the MacAdam Student Observatory (MSO) on the University of Kentucky campus in Lexington, Kentucky, the University of Louisville's Moore Observatory (ULMO) near Louisville, Kentucky, and in one case, the University of Southern Queensland's Mt. Kent Observatory, near the town of Toowoomba in eastern Australia.\par

All three telescopes were 20-inch PlaneWave Corrected Dall-Kirkham f/6.8 models (CDK20). Filters used were $V$, $G$, and in one case, $R$. The Moore data was reduced using \textit{AstroImageJ} \citep{collins2016astroimagej}. Fourth-order polynomials were fit to the relevant parts of the light curves in order to determine an epoch of maximum in JD and calculate the phases of the spectroscopy. 

The photometry for a given star from MSO, LMSO, and Mt. Kent were from a single night to capture the maximum light, with a dense sampling cadence of $\approx$ 20 sec to 2.5 min, so as to establish the time at which the pulsation phase was zero. Pixel scales at these telescopes were 
$\approx$ 0.4 -- 1''/pixel.

To make the most precise determinations of phases following acquisition of the stars' spectra, we also obtained photometry from the Kilodegree Extremely Little Telescope (KELT) and the space-based Transiting Exoplanet Survey Satellite (TESS) \citep{pepper2007kilodegree,ricker2015}.  The KELT and TESS observations use wide fields of view of 26$^{\circ}$ × 26$^{\circ}$ and 24$^{\circ}$ × 24$^{\circ}$ from a single camera, respectively, larger pixel scales of $\sim$23''/pixel and $\sim$21''/pixel, and observing baselines stretching over months or years, at typical sampling cadences during observing windows of $\approx$ 2-30 min \citep{pepper2012kelt}. The long temporal baselines of KELT and TESS, and the extra benefit of the excellent photometric precision of TESS, enabled the dense phase sampling required for establishing highly precise pulsation periods.

\setcounter{table}{0}
\begin{deluxetable*}{lllll}
\tablecaption{Photometric Periods of Program Stars}
\tablecolumns{5}
\tabletypesize{\footnotesize}
\tablehead{Star & Type & Photometry source & Photometry source & Period (day)$^{a}$ \\
&  & (for epoch-of-max) & (for period) & 
}
\startdata
RW Ari$^{c}$ 		& c 		& MSO 	&  TESS  			& 0.35431(12)*	  \\
X Ari	 			& ab    	& MSO 	& TESS 			& 0.6511808(76)	  \\
UY Cam				& c 		& ULMO & TESS  			& 0.26703933(47) \\
RR Cet				& ab 	& MSO 	& TESS  			& 0.553037(16)  \\
SV Eri				& ab 	& Mt. Kent		& TESS  & 	0.7138091(19)  \\
VX Her$^{b}$		& ab 	& ULMO & AAVSO  		& 	0.45535897(53) 	        \\ 
RR Leo				& ab 	& ULMO & TESS, KELT   	& 	0.4524032(11)  \\ 
TT Lyn				& ab 	& MSO 	& TESS, KELT  	& 0.5974318(41) \\ 
TV Lyn				& c 		& ULMO & TESS, KELT  	& 0.2406505(31)    \\ 
TW Lyn				& ab 	& MSO 	& TESS, KELT  	& 	 	0.4818637(80)        \\ 
RR Lyr$^{b}$				& ab 	& MSO 	& TESS, KELT  	& 	 	0.566829(41)      \\ 
V535 Mon$^{b}$		& c 		& KELT & KELT  			& 0.3258(36)*	  \\
V445 Oph			& ab 	& MSO 	& KELT  			& 	 0.39702436(20)        \\ 
AV Peg				& ab 	& ULMO & KELT 			& 	 0.39038168(88)        \\ 
BH Peg$^{b}$				& ab 	& ULMO & \citet{monson2017standard} &	 0.640993(34)	        \\
AR Per				& ab 	& ULMO & KELT  			& 	 0.42555047(39)        \\ 
RU Psc$^{b,c}$		& c 		& MSO 	& KELT  			&  0.3902477(15)* 	        \\
T Sex				& c 		& MSO 	& TESS  			& 0.32474752(28)   \\
TU UMa				& ab 	& MSO 	&  TESS, KELT 	& 	 0.557654(12)	        \\ 
\enddata
\tablecomments{
$^a$~For compactness of notation, the uncertainties in parentheses correspond to the last two significant digits (e.g., 0.354311(17) means 0.354311$\pm$1.7$\times 10^{-5}$).\\
$^b$~Blazhko star (e.g., \citealt{skarka2013known})\\
$^c$~Period is known to vary \\
*~The period of this star is poorly defined over long time baselines. The value tabulated here is based on photometry taken close in time to when spectra of the star were obtained at McDonald Observatory.
}
\label{table:star_periods}
\end{deluxetable*}

\subsection{Spectroscopy}
\label{subsec:spectroscopy}
 
The spectroscopy was gathered with 2.08 m Otto Struve Telescope at McDonald Observatory in Texas, using the low-to-moderate resolution Electronic Spectrograph Number 2 (ES2), with $R\sim1,000$. The two observing runs took place on UT 2012 December 21-28, and UT 2013 July 20-26.\par

\begin{figure}
\centering

\includegraphics[width=1.0\linewidth]{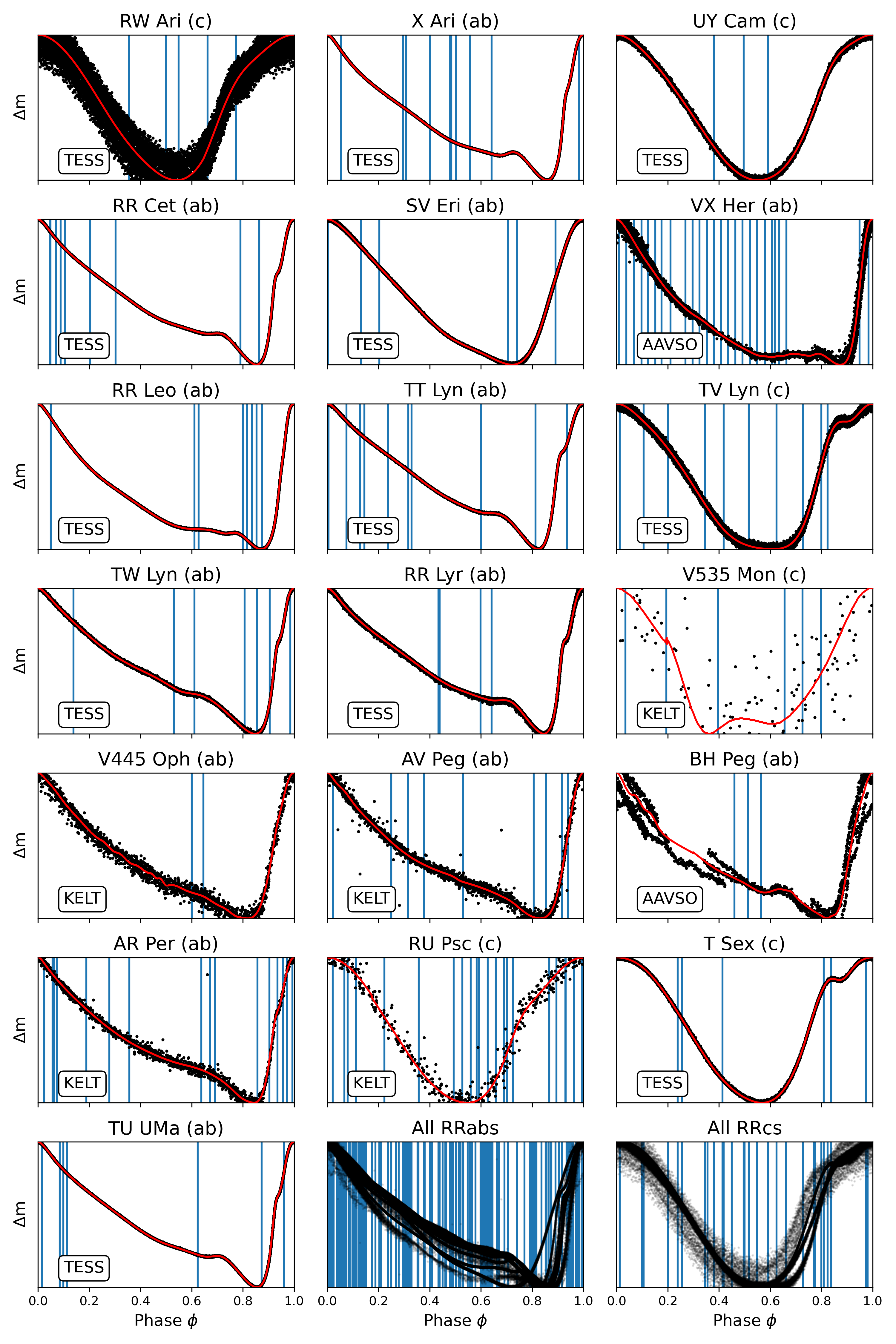}
\caption{The phase-folded light curves of our program stars, in the style of Fig.\ 1 of \citet{pancino2015chemical}. Red lines are spline fits from the folding process, and vertical blue lines indicate the phase epochs of our spectra. Vertical axes are scaled to fit the spline fits, leaving some photometric points outside the display region, so as to make it easier to see the curves. RU Psc, RW Ari, and V535 Mon have poorly defined periods (e.g., \citealt{wils2008recent}; A. Odell, pers. comm.). Photometry for those stars were limited to short baselines at times near or around the time the spectra were taken. Offsets in the BH Peg subplot may be due to astrophysical Blazhko changes (period $\sim$10s to $\sim$100 days; \citealt{smith1995,skarka2014bright}) over the 6-year span of the AAVSO observations, or changing photometric zero points in the observations. Overlaid plots at the bottom centre and right exclude RU Psc and V535 Mon. (See Appendix \ref{sec:secFitting} and Sec.\  \ref{subsec:spectroscopy}.)} 
\label{fig:pancino_style}
\end{figure}

All told, 13 stars were observed at McDonald Observatory in December 2012, and 6 in July 2013, totaling 18 different stars, including RU Psc, which was observed during both runs. Of these stars, 13 were RRab stars. A total of 169 useable spectra were reduced in the analysis (see Fig.\ \ref{fig:pancino_style}). The stars whose periods were covered as much as possible were VX Her (an RRab, with 26 spectra) and RU Psc (an RRc, with 22 spectra).\par

Raw spectra from December 2012 ranged over wavelengths of 3900\,{\AA} to 5000\,{\AA}, and in July 2013, 3610\,{\AA} to 4960\,{\AA}. Spectra were later truncated to the range 3911\,{\AA} to 4950\,{\AA}, to limit distortions of the Ca II K and H$\beta$ absorption lines during the normalization process. Integration times ranged from 40 to 1800 seconds, and always remained less than 10\% of the period in order to avoid smearing of the phase. Of the 169 spectra, the integration times of six of them represented less than 1\% of the period, 52 represented between 1\% and 2\%, 86 between 2\% and 5\%, and 24 between 5\% and 9\%. In a few instances, integration was paused due to a passing cloud, which may have increased the total elapsed time by a minute or two. 

\subsection{Calculation of Periods and Phases}

After the spectroscopy were gathered at McDonald Observatory, we re-calculated periods for most of the stars using phase-folded photometry from KELT and TESS. The heavy sampling of photometry in these long-term surveys maximizes the precision of the periods, which minimizes the error in the calculated phase based on a given epoch-of-maximum. Two of our program stars were unavailable in KELT or TESS: for VX Her we obtained photometry from the AAVSO, and for BH Peg we used the period in \citet{monson2017standard}. 

The KELT light curves used dates in terrestrial time, so the dates were converted to barycentric Julian dates (BJD) using \texttt{astropy} \citep{Astropy2013A&A,Astropy2018AJ}. TESS light curves were downloaded from the online Mikulski Archive for
Space Telescopes (MAST), and were already in BJD.

\begin{figure}
\centering

\includegraphics[trim=0cm 0cm 0cm 0cm,width=1.0\linewidth,clip=True]{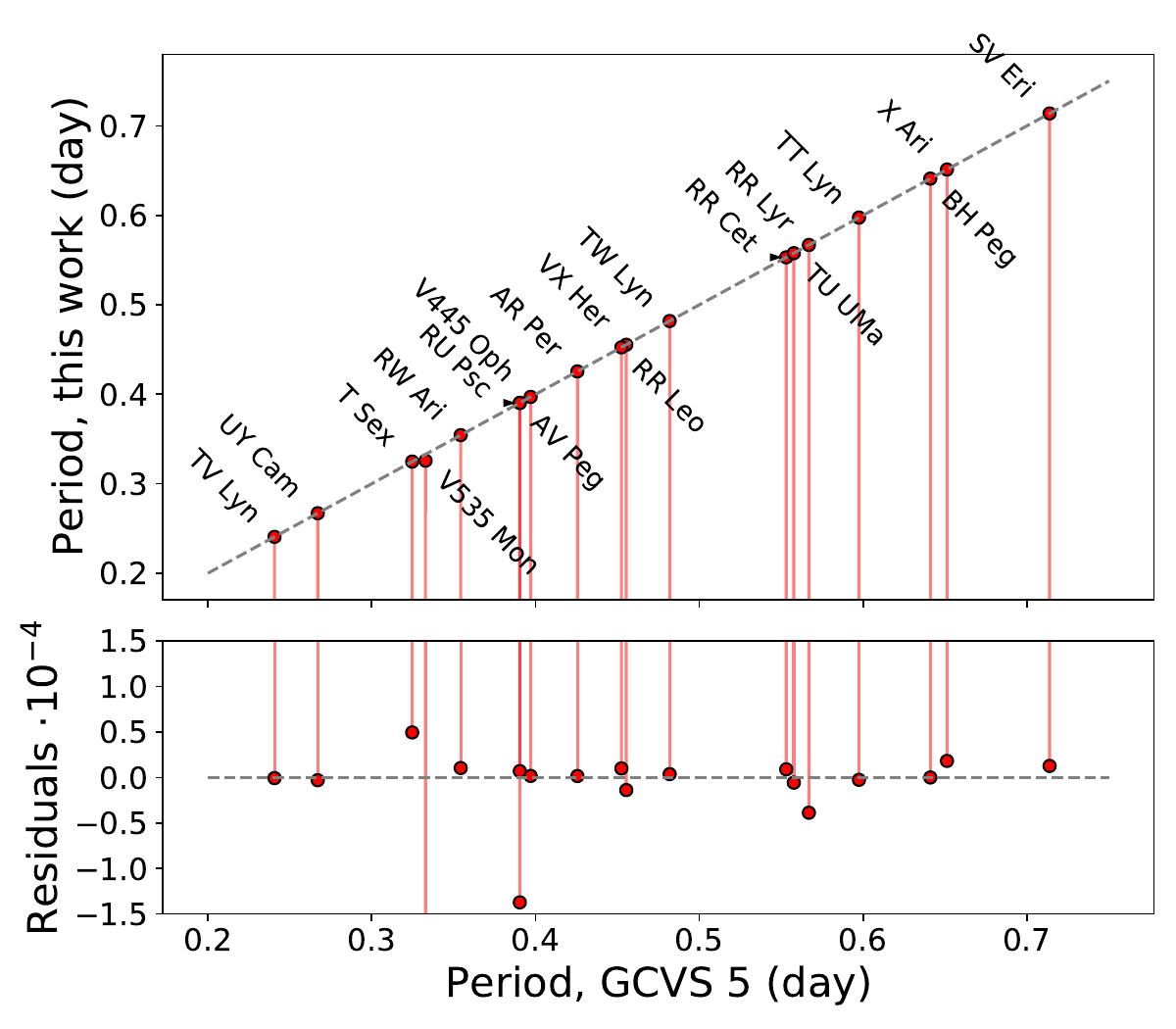}
\caption{Periods we use for our program stars, compared to those in the GCVS 5 catalog of variable stars. The dashed line is one-to-one in the top plot, and zero in the bottom plot. Vertical lines connect the same star to guide the eye. The bottom plot is in units of $10^{-4}$ day, or 8.64 sec. The residual of the period of V535 Mon is outside the plot, at $-70.734\times 10^{-4}$ day = 611 sec.}
\label{fig:period_comparison}
\end{figure}

The \texttt{VARTOOLS} light-curve analysis program \citep{hartman2016vartools} was used to provide a common interface as we tried several different period-finding algorithms. We chose to use the Fast-$\chi^2$ algorithm \citep{palmer2009fchi2}, limiting our search to between 0.1 and 0.9 days, with the number of harmonics set to 3 (fundamental plus the first two harmonics). The periodogram was over-sampled by a factor of 10 and the two most significant peaks were analyzed. For many stars in the KELT footprint, separate curves taken from both east and west of the meridian were available. In such a case, the period we adopted for the rest of the analysis corresponded to the best periodogram peak in the 
best-fitting light curve (East or West) as determined by eye. We compared these to the published periods from the General Catalog of Variable Stars (GCVS, \citealt{samusetal2017}) to check for consistency (Fig. \ref{fig:period_comparison}). 

For BH Peg, we used the period as determined by \citet{monson2017standard}, who used archival data with a baseline of thirty years, together with robotic observations from the Three-hundred MilliMeter Telescope (TMMT) at Las Campanas Observatory. The data were phase folded across multiple basis wavelengths, and transformed onto a basis filter set of Johnson $B$, $V$, Kron-Cousins $I_{C}$, 2MASS $J$, $H$, $K_{S}$, and Spitzer [3.6] and [4.5] $\mu$m channels. We used the period from \citet{monson2017standard} to phasefold the AAVSO photometry, which we limited for self consistency to observations contributed from 2015 to 2021 by one observer, G. Samolyk. We found that the folded light curve was the most consistent of any produced with other literature period values, or period values found from our own phase folding of sparsely sampled AAVSO data.

It became clear that not all of our program stars have well-defined periods. The KELT light curves for RU Psc and V535 Mon showed significant variations in the pulsation period, the envelope amplitude of the pulsations characteristic of Blazkho variations, or both, to the extent that they could not be phased with the entirety of the available photometry. We therefore restricted the photometry to a short baseline, closest in time to when the spectroscopy was collected. For RU Psc we used the East KELT light curve limited to BJDs 2456187.75 to 2456663.58. For V535 Mon we limited the West light curve to BJDs 2456281.76 to 2456304.98. In addition, RW Ari is known to have an irregular pulsation (A. Odell, pers. comm.).

To determine the epoch of maximum brightness, we used the derived periods to phase the light-curve data. The phased light curves were then binned into 100 phase bins and the median of each bin was determined. In the cases of V535 Mon, RU Psc, and BH Peg, 20, 50, and 65 bins, respectively, were used due to a smaller number of data points. The binned data was then fit with a 3rd order univariate spline from the SciPy \texttt{interpolate} package, and the brightest point on the spline curve was taken to be the maximum. The phase of maximum was then converted back into an epoch and used to re-phase the light curve, so that maximum light occurred at zero phase. The periods of our program stars were used to refine the corresponding phases of the spectra from McDonald Observatory, based on an epoch of maximum brightness chosen to be close in time to the spectroscopy. The sources of photometry and final periods are listed in Table \ref{table:star_periods}.

\section{Data Analysis}
\label{sec:secAnalysis}

\subsection{Reduction}

\subsubsection{Reduction of synthetic spectra}
\label{subsubsec:syntheticSpectra}

We wrote a pipeline in Python to take the synthetic spectra and generate a higher-order version of the calibration of  \citetalias{layden1994}, using the Python version of \texttt{Robospect} to measure EWs \citep{watersetal2013}. 

Our expanded version of the \citetalias{layden1994} relation is

\begin{equation} \label{eqn:expanded_lay_text}
\begin{split}
W(K)= a&+b\,\cdot W(H)+c\, \cdot [\textrm{Fe/H}] \\
&+d\,\cdot W(H)\cdot [\textrm{Fe/H}]+f\,\cdot W(H)^{2}\\
&+g\,\cdot [\textrm{Fe/H}]^{2}+h\,\cdot [\textrm{Fe/H}]\cdot W(H)^{2}\\
&+k\,\cdot W(H)\cdot [\textrm{Fe/H}]^{2}\\
\end{split}
\end{equation}

\noindent
where $W(K)$ is the EW of the Ca II K line and $W(H)$ is the EW in angstroms of the Balmer lines. Coefficients $\{a,b,c,d\}$ are used in \citetalias{layden1994}, and our addition of $\{f,g,h,k\}$ is explained in the Appendix \ref{sec:secFitting}.

\begin{figure}
\centering
\includegraphics[width=1.0\linewidth]{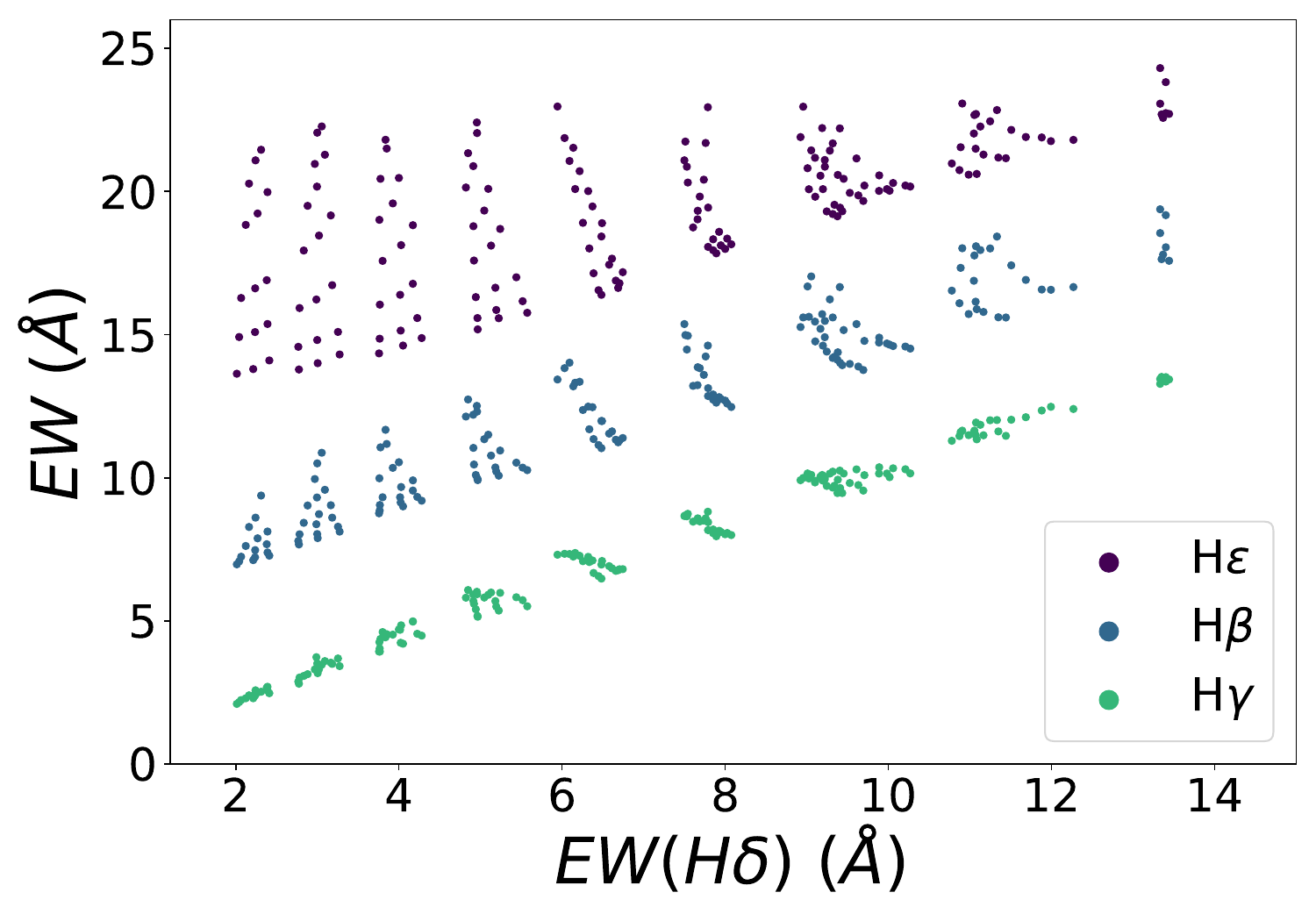}
\caption{A comparison of the EWs of hydrogen lines with that of H$\delta$ (filled data points), showing the strong linear trend between H$_{\gamma}$ and H$_{\delta}$. For clarity, H$_{\epsilon}$ and H$_{\beta}$ have been offset, and error bars have been removed.} 
\label{fig:balmer_comparisons}
\end{figure}

The H$\beta$ line EWs were discarded from the analysis because they introduced considerable scatter in both the synthetic and empirical spectra, apparently due to metal contamination, as was also noted by \citetalias{layden1994} (Fig.\ \ref{fig:balmer_comparisons}).

We checked for consistency in the EWs returned by \texttt{Robospect} and manual use of \texttt{IRAF} \citep{iraf_1986}.\footnote{IRAF was distributed by the National Optical Astronomy Observatory, which was managed by the Association of Universities for Research in Astronomy (AURA) under a cooperative agreement with the National Science Foundation.} With \texttt{IRAF} we tested window sizes of 10, 28, and 56\,{\AA} on the synthetic spectra, and found results between \texttt{Robospect} and \texttt{IRAF} to be consistent with window sizes set to 56\,{\AA}. 

The steps of our pipeline workflow are as follows: 

\begin{enumerate}
\item Normalize the spectra
\item Measure EWs and errors for Ca II K, H$\delta$, and H$\gamma$ using \texttt{Robospect}
\item Generate a net Balmer EW from a simple average between the H$\gamma$ and H$\delta$ EWs
\item Feed the EWs, errors, and input metallicities into an MCMC that samples the posteriors of the coefficients and their errors as per the expanded version of the relation in \citetalias{layden1994}, using the \texttt{emcee} package \citep{emcee} 
%\item Map the observed changes in the Ca II K and Balmer EWs of the star as found in step 5. (Table \ref{table:sources}).
\item Save the sampled posteriors for application to empirical spectra
\end{enumerate}

\subsubsection{Reduction of empirical spectra}

CCD readouts of the spectra taken at McDonald Observatory were reduced with \texttt{IRAF}. Finding a solution to the shorter-wavelength end of the spectrum proved challenging, due to the presence of no more than two reliable calibration-lamp emission lines at wavelengths less than about 4150\,{\AA} (which includes the H$\delta$ and Ca II K absorption lines in the object spectra), and a complete lack of lines below 3950\,{\AA} (which includes Ca II K).

\begin{figure*}
\centering
\includegraphics[width=1.0\linewidth]{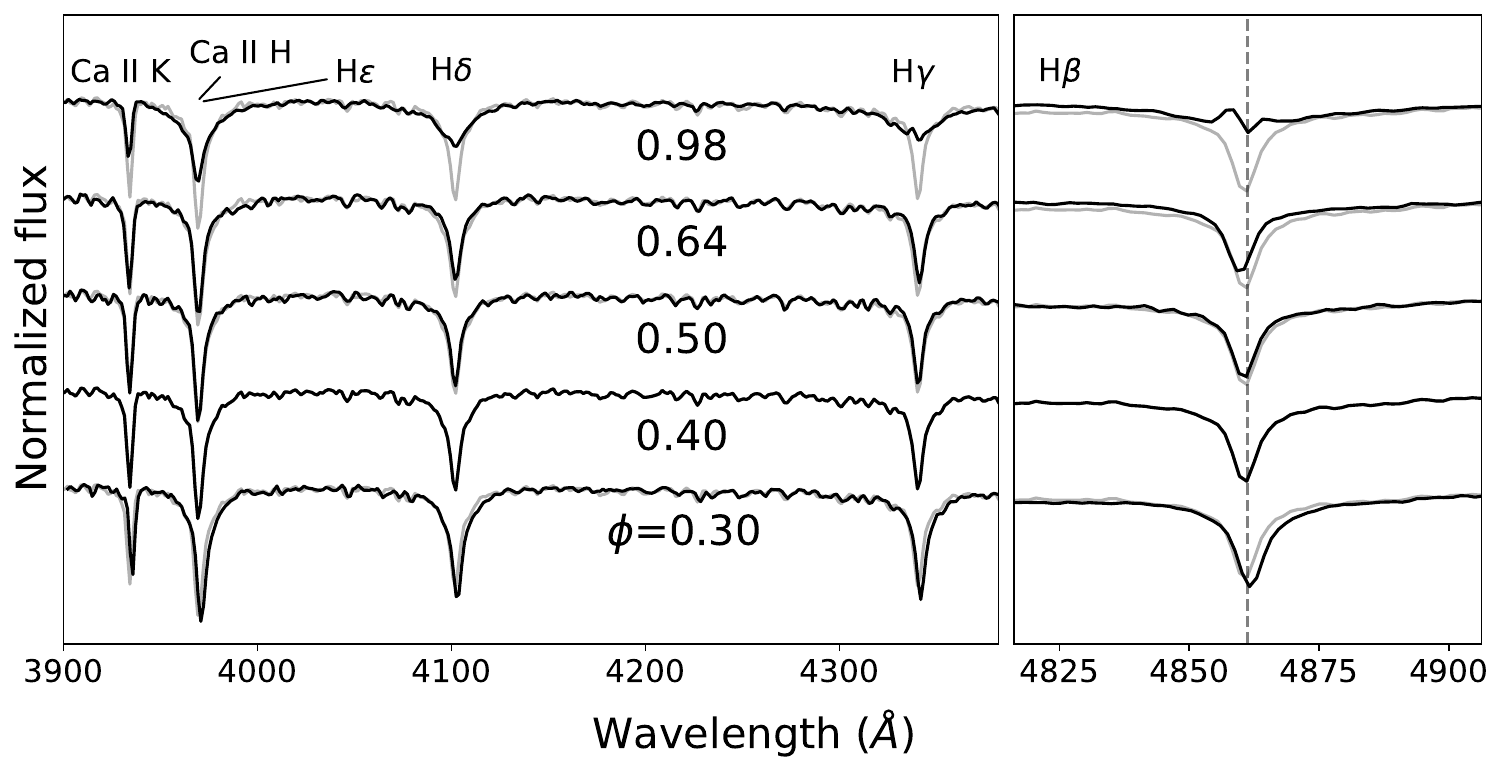}
\caption{Normalized spectra of X Ari at different phases. Spectra are offset for clarity, with the spectrum at phase 0.4 underplotted in grey for comparison. The pulsation shock at late phases leads to a filling-in of lines.}
\label{fig:x_ari_shock}
\end{figure*}

Spectra were written out to ASCII files, and cosmic rays were manually removed by deleting data lines in the ASCII file. Fig. \ref{fig:x_ari_shock} shows example spectra taken of X Ari, including the time at which a shock travels through the photosphere. For each star, we assigned the metallicity found from the mapping of [Fe/H] from high-resolution studies as described in Sec.\  \ref{subsec:lit_metal}.

\subsection{Contamination from Interstellar Calcium}
\label{subsec:subsec_contamination}

The presence of interstellar calcium along the line-of-sight to an observed RR Lyrae can add to the total EW of the measured Ca II K line, and systematically increase the star's computed metal abundance.  To compensate for this contamination, \citetalias{layden1994} applied the interstellar calcium EW model of \citet{beers1990estimation}, which is a function of a star's Galactic latitude and vertical height above the Galactic plane. This model allowed for the subtraction of the EW of the interstellar component from the EW of the net Ca II K line. 

However, SDSS spectroscopy since then has revealed highly textured interstellar Ca II K absorption \citep{murga2015calcium} within a large observational footprint that contains some, but not all, of our program stars. Even if the line-of-sight ISM calcium absorption to a single star can be quantified precisely---and the resolution of the map in \citet{murga2015calcium} is about 3.7$^{\circ}\times$3.7$^{\circ}$---the saturated Ca II K stellar line will experience shifts of up to $\approx$ 70 km s$^{-1}$, in addition to the intrinsic radial velocity of the star. Thus, the ISM will contaminate the stellar Ca II K line different amounts at different times.

\begin{figure}
\centering
\includegraphics[trim=1.2cm 0cm 0cm 0cm,clip,width=1\linewidth]{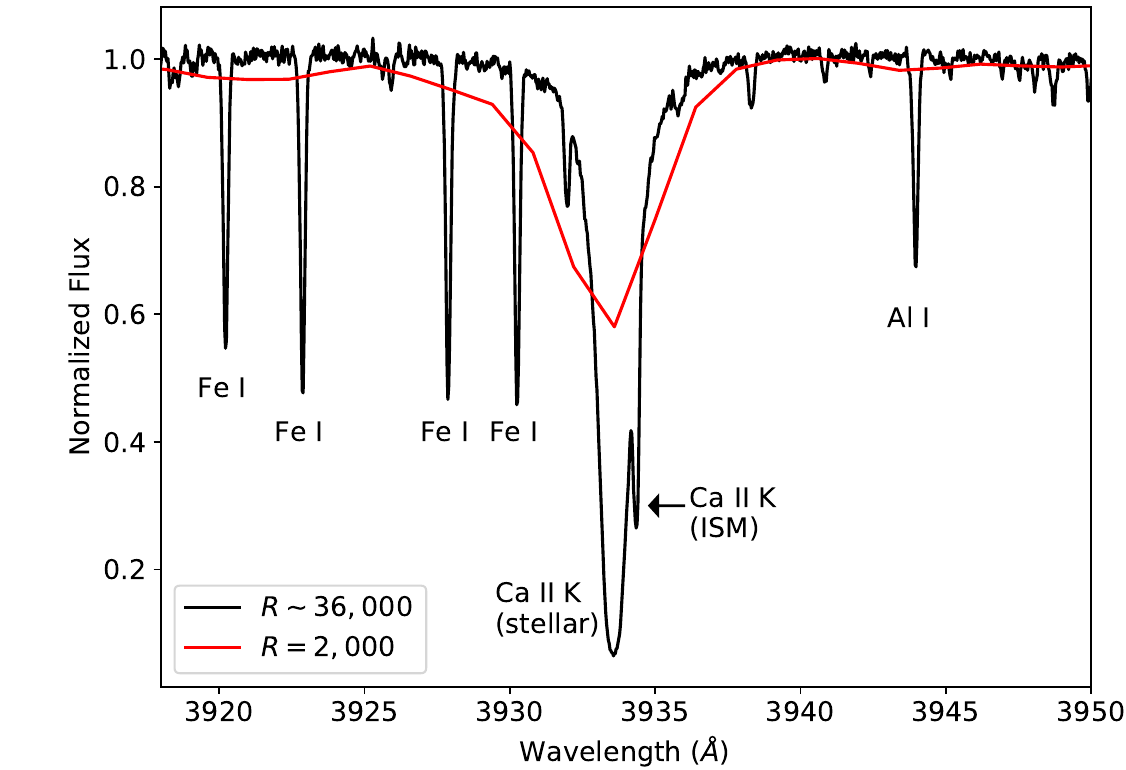}
\caption{High-resolution spectroscopy from \citet{nemecetal2013} of the metal-poor star X Ari. Lines are identified with the model atmosphere program \texttt{SPECTRUM} \citep{gray_model}. The fortuitous pulsation phase of the star allows separation between the stellar and interstellar Ca II K line. Data kindly provided by J. Nemec. Also see Sec.\  \ref{subsec:subsec_contamination}. 
}
\label{fig:nemec_highres}
\end{figure}

To see whether it is worth modeling these effects, we investigated the effect of the interstellar calcium on our most metal-poor program star, X Ari.  The interstellar Ca II K line is readily seen redward of its stellar counterpart in the high-resolution spectrum in Fig.\ \ref{fig:nemec_highres}. At this resolution, the two lines can easily be deblended.  When using low-resolution spectroscopy ($R\approx 2000$), however, the contribution of the interstellar component is blended into the line and cannot be directly removed.  

We measured the EW of the interstellar Ca II K line in the high-resolution spectroscopy from \citet{nemecetal2013} to be 80\,m{\AA}.  Inclusion of this contaminant increases the net Ca II K EW by 4.6\%.  After smoothing the spectral resolution to that of $R=2000$, the increase of the EW (which includes the interstellar component) as measured with a Gaussian fit, over the high-resolution, deblended stellar EW, is 1.2\%.  This is well below the random fitting error of the low-resolution line. 

\begin{figure}
\centering
\includegraphics[trim=0cm 0cm 0cm 0cm,clip,width=1\linewidth]{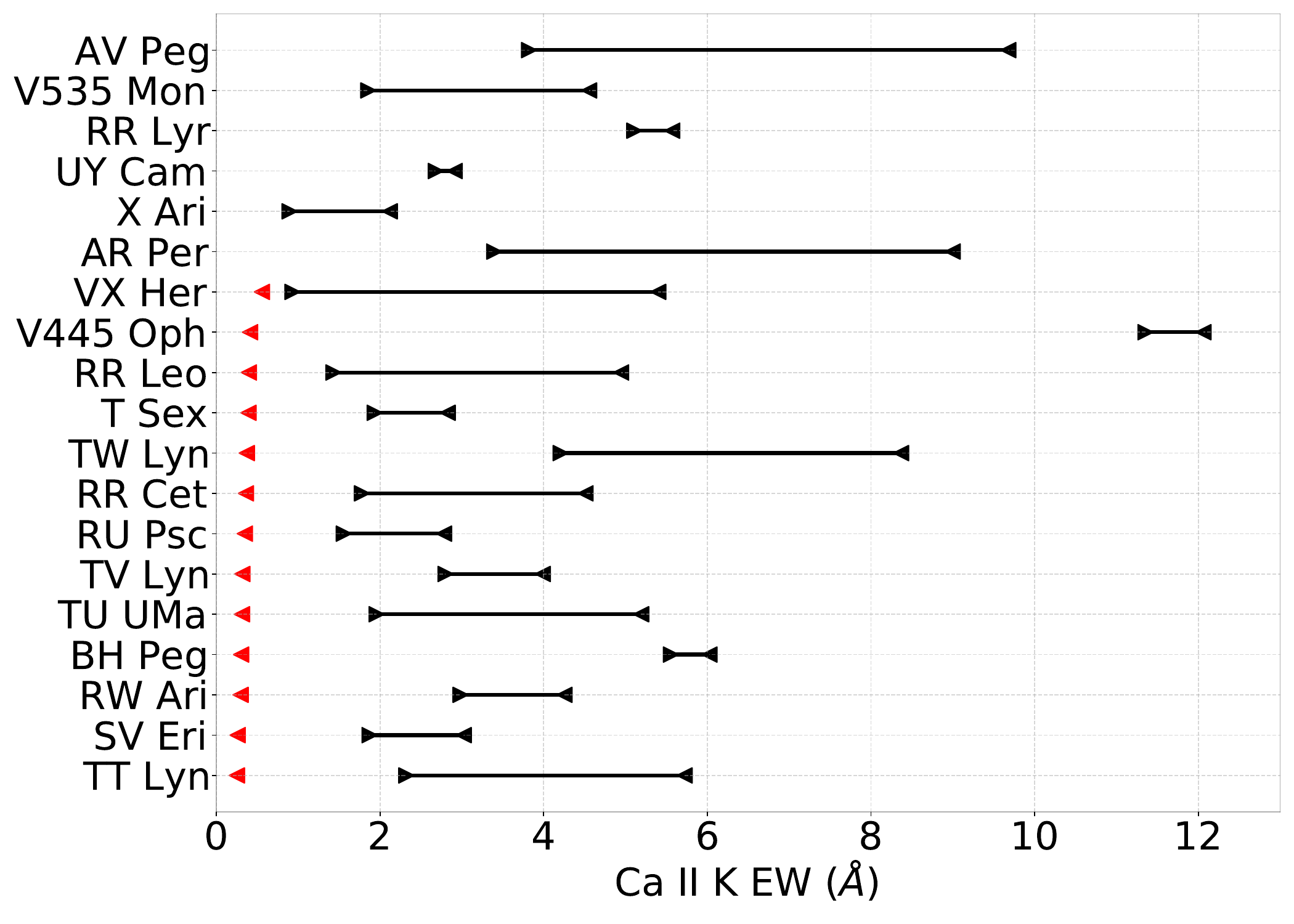}
\caption{Comparison of the smallest and largest Ca II K EWs measured from our program star spectra (black), and ISM Ca absorption as mapped by \citet{murga2015calcium} (red). Program stars at the top are not within 7.4$^{\circ}$ of the footprint of \citet{murga2015calcium}.}
\label{fig:murga_comparison}
\end{figure}

We also used the maps of \citet{murga2015calcium} to find upper bounds to ISM Ca II K absorption close to the lines-of-sight towards our other program stars. The maps were generated from the integrated absorption along lines-of-sight to distant quasars and galaxies, and thus represent an over-estimate of the absorption to our stars, which are at 0.25 kpc $<d<$ 2.1 kpc \citep{gaia_edr3}. Fig.\ \ref{fig:murga_comparison} compares Ca II K EWs of our stars with values of ISM absorption, which we take to be the maximum of the ISM absorption within an angle of 7.4$^{\circ}$ (two resolution elements) of the star. For the stars which appear in or near the footprint, the upper bounds of the ISM are never larger than the smallest Ca II K EWs, and are an average of 0.18\,{\AA} of the minimum Ca II K EW for each star, and 0.08\,{\AA} of the maximum.

Based on the high-resolution spectrum of X Ari and the upper bounds on the ISM absorption, we conclude that the measured Ca II K EWs of RR Lyrae variables with $\textrm{[Fe/H]} > -2.5$ will be essentially unaffected by interstellar calcium, due to the significantly larger EW of the stellar line, regardless of pulsation phase. The overall small and variable contribution to the stellar EW as measured from low-resolution spectroscopy justifies our decision to not attempt to correct for the interstellar component in our calibration.

\begin{figure*}
\centering
\includegraphics[trim=0cm 0cm 16cm 0cm,clip=True,width=1.0\linewidth]{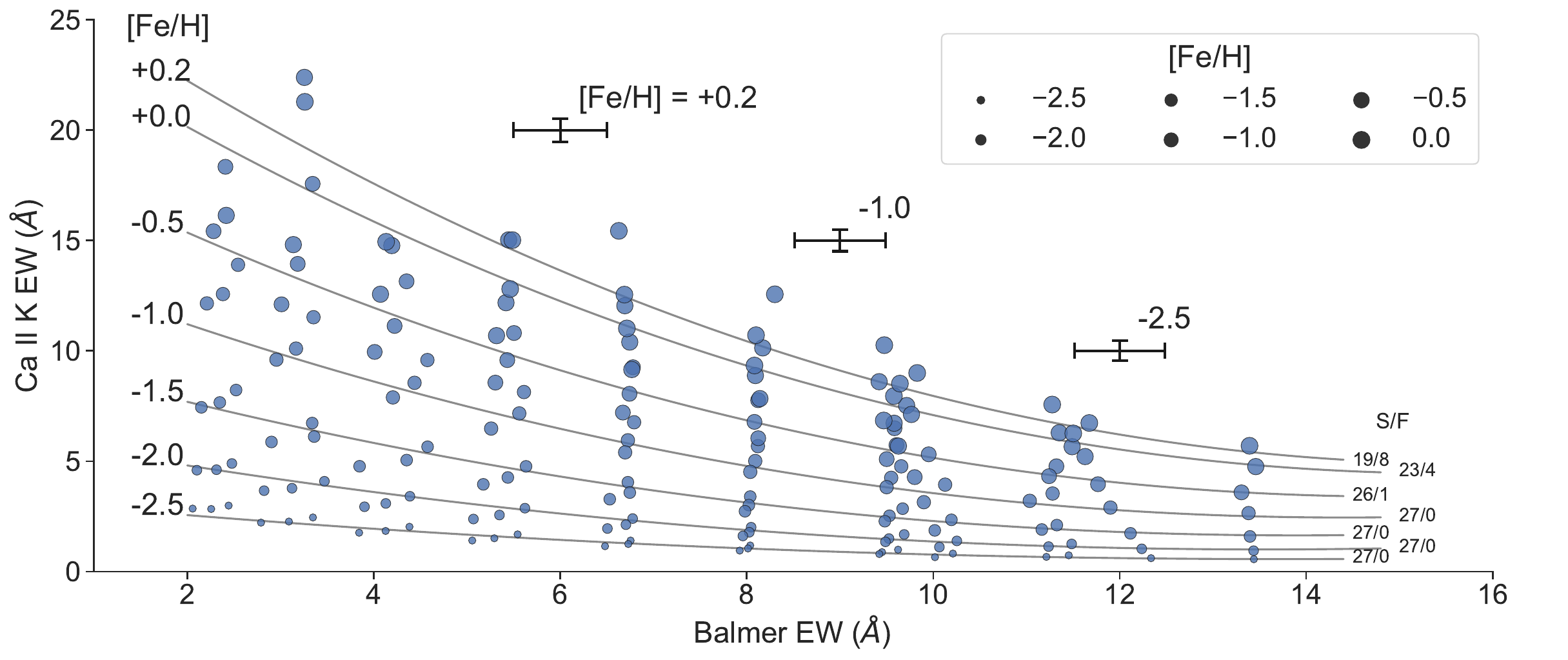}
\caption{The space of Ca II K and Balmer EWs of the synthetic spectra, with iso-metallicity contours from the final calibration. Points with error bars show the median sizes of the error bars at selected metallicities. The S/F ratio indicates the number of spectra with successful and failed line fits. 
}
\label{fig:kh_synth}
\end{figure*}

\section{Results}
\label{sec:secResults}

\subsection{Coefficients}
\label{subsec:coeffs}

Fig.\ \ref{fig:kh_synth} shows a plot of the measured EWs from the synthetic spectra and iso-metallicity contours from the fit. The original \citetalias{layden1994} calibration implied a Ca II K EW of zero at a Balmer-line EW of $\approx$ 12\,{\AA} for [Fe/H] = $-1$. Our calibration is comparatively `shallower,' and can handle a larger range of Balmer-line EWs. 

Errors in the EWs as measured by \texttt{Robospect} were larger than would be expected based on the scatter evident in Fig.\ \ref{fig:kh_synth}, so we fit a linear scaling relation between the \texttt{Robospect} errors and the standard deviation in measured EWs from the star VX Her, for which we obtained particularly heavy phase coverage with our spectroscopy (Fig.\ \ref{fig:pancino_style}). We subdivided the spectra of VX Her into phase bins of $\Delta \phi=0.25$, found the standard deviations therein, and compared it to the mean of  \texttt{Robospect} errors for the same lines. Based on this exercise, we implement in our pipeline a rescaling of the errors returned by \texttt{Robospect}, and consider this the true error in EW.

\subsection{Tests on McDonald Observatory Data}
\label{subsec:mcd_app}

\begin{figure*}
\centering
\includegraphics[trim=0cm 0cm 17cm 0cm,clip=True,width=1.0\linewidth]{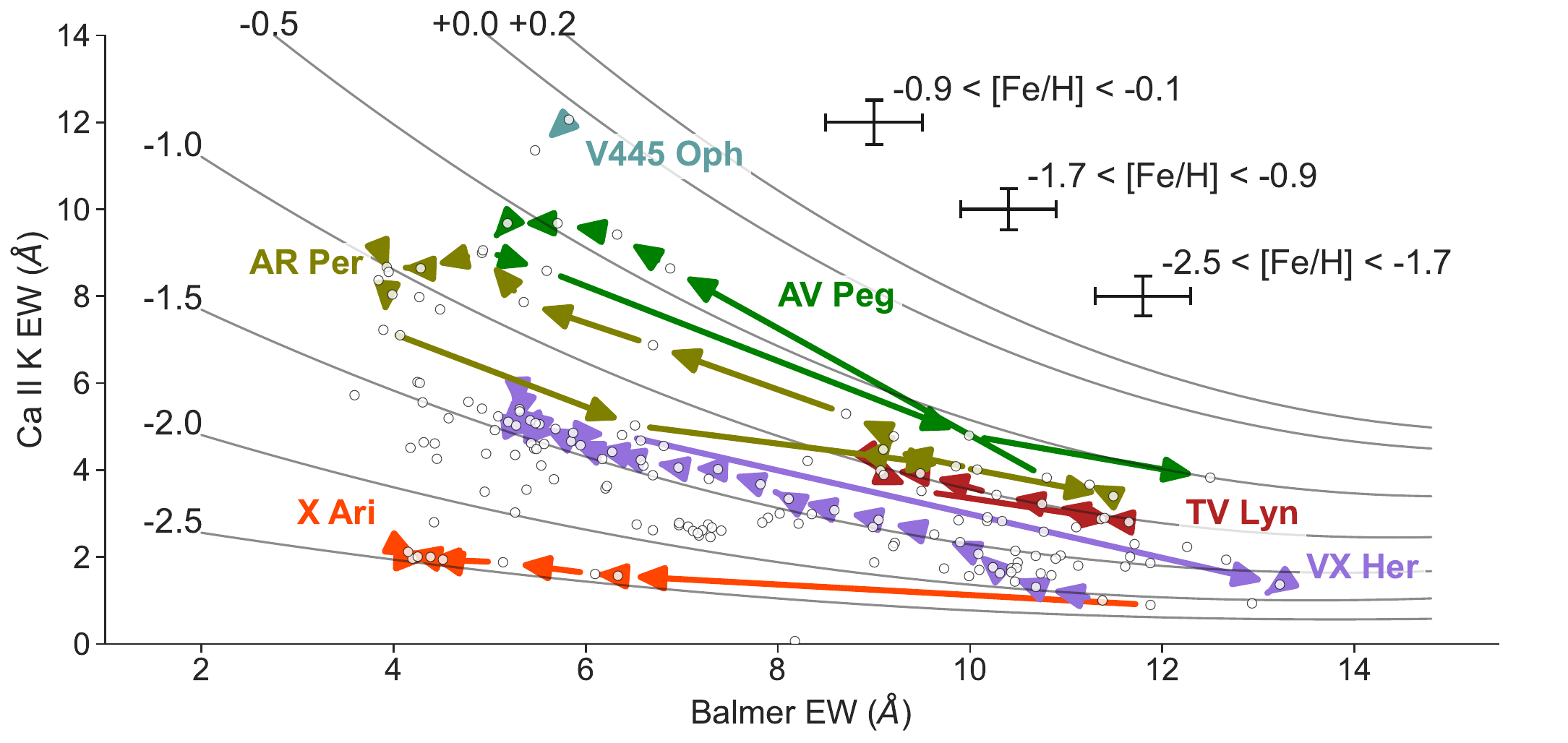}
\caption{The space of Ca II K and Balmer EWs for the program stars, with the same contours and relative [Fe/H] marker size scale as in Fig.\ \ref{fig:kh_synth}. The ``looping'' cycle in the plane is illustrated for six stars, where phase increases along the direction of the arrows. 
}
\label{fig:kh_mcd}
\end{figure*}

\begin{figure}
\centering
\includegraphics[trim={0cm, 18cm 0cm, 0cm}, clip=True, width=1.0\linewidth]{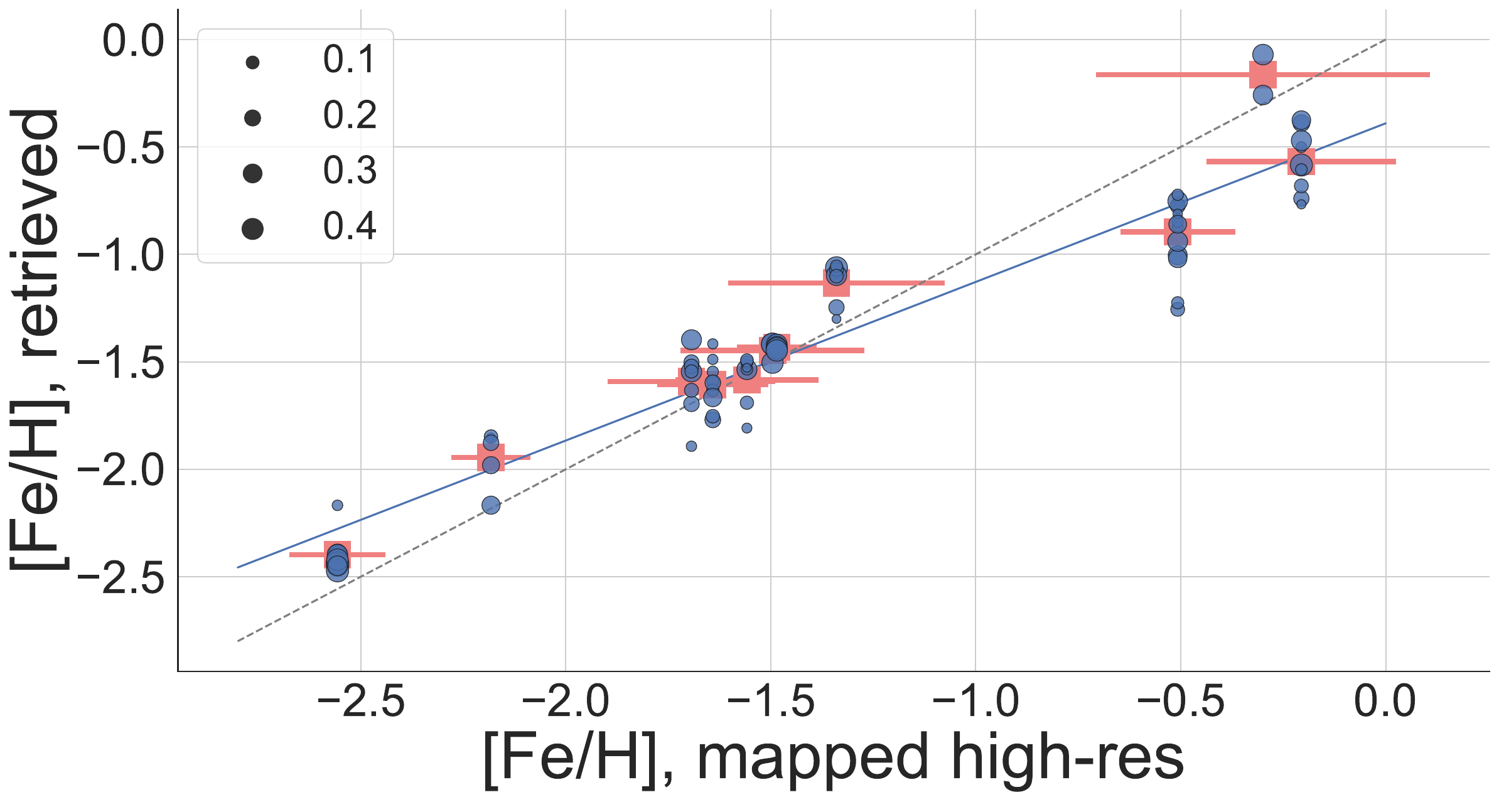}
\caption{Program RRab star [Fe/H]s as estimated by our calibration, compared to their values based on high-resolution spectroscopy. Each data point corresponds to one spectrum. The size of the dots indicate the absolute value of distance in phase from the maximum of the light curve (i.e., $\phi=\{0.2,0.7\}$ maps to $|\Delta\phi|=\{0.2,0.3\}$. The dashed line is one-to-one or zero, and the solid line is the best fit. Red squares are average values taken across all the spectra for that star, and are the same as the retrieved values listed in the appendix in Table \ref{table:prog_star_fehs}.}
\label{fig:us_highres_mcd_stars}
\end{figure}

This calibration was tested on McDonald Observatory spectra of both RRab and RRc stars. Out of 169 spectra, 9 had line-fit failures. Fig.\ \ref{fig:kh_mcd} shows the measured EWs, and Fig.\ \ref{fig:us_highres_mcd_stars} shows the results of the metallicity estimates, compared to what one would expect from high-resolution studies. After visual inspection, the bad-phase region was found to be so narrow in phase given our sampling that it was disregarded. 

There is an over-estimation of [Fe/H] at the metal-poor end, and an under-estimation at the metal-rich end. To counteract this, we implemented a final correction in the pipeline to effectively map the best-fitting  line onto the one-to-one line. 

\subsection{Tests on SDSS Spectra}
\label{subsec:stripe82}

This calibration was tested on single-epoch SDSS spectra of unique stars that appear in the footprint of the SDSS Stripe 82 or the photometric Catalina Sky Survey \citep{larson2003css,drake2013catalina,abbas2014}. Stripe 82 is a subsection of the SDSS footprint along the celestial equator that was imaged 70 to 90 times between 1998 and 2007 \citep{jiang2014sloan}. After manually removing cosmic rays, Stripe 82 spectra were processed in the same way as the McDonald spectra.  \par

\begin{figure}
\centering
\includegraphics[width=1.0\linewidth]{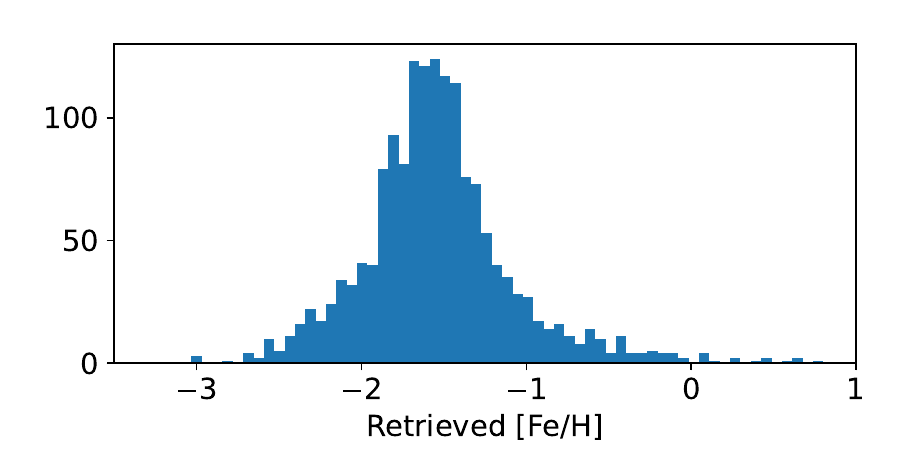}
\caption{Histogram of estimated [Fe/H] of RR Lyrae stars in Stripe 82 and the Catalina footprint.}
\label{fig:sdss_test}
\end{figure}

\begin{figure}[!htb]
\centering
\includegraphics[trim=0.0cm 0.0cm 0.3cm 0.2cm,clip=True,width=1.\linewidth]{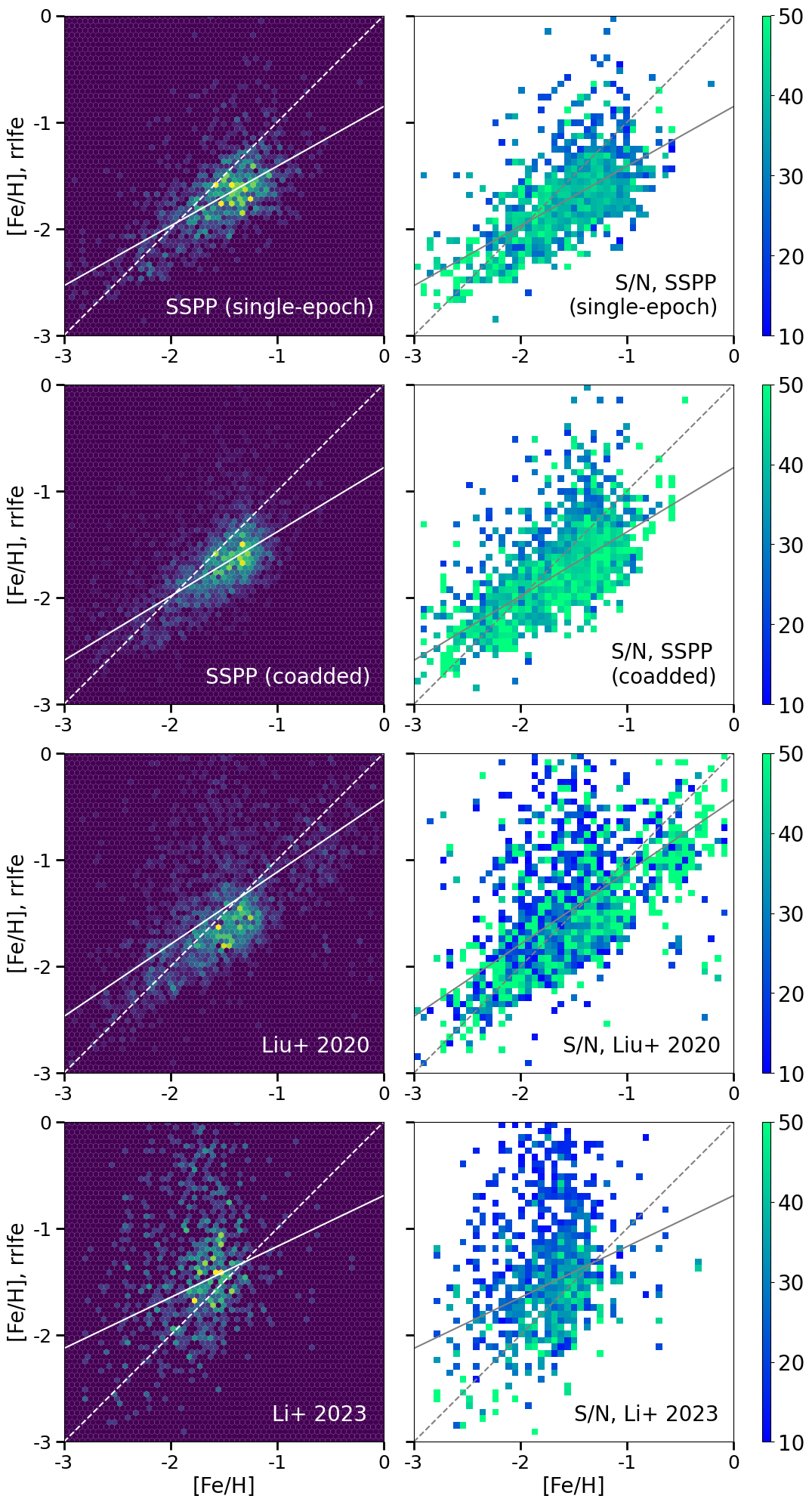}
\caption{Estimates of [Fe/H] of RR Lyraes from SDSS spectra, as compared with the SSPP pipeline, \citet{liu2020probing}, and \citet{lietal2023}. Left column: histograms; right column: binned S/N of the SDSS spectra upon which our estimates are based. S/N$\geq$10 for all spectra, and the scale is truncated at 50. Dashed lines are one-to-one, and solid lines are best fits weighted by S/N. The SSPP values are from spectra acquired separately, or coadded into one, with no phase information. \citet{liu2020probing} used synthetic spectral template matching to estimate [Fe/H]. When more than one spectrum was available for a star they used observed spectra at phases unaffected by atmospheric shocks, based on the EW of the Ca II K line. \citet{lietal2023} used photometric relations based on the metallicities of \citet{liu2020probing}.}
\label{fig:comparison_retrievals_others}
\end{figure}

Fig.\ \ref{fig:sdss_test} shows that the SDSS spectra have a clear peak in at metallicity $\textrm{[Fe/H]}\approx -1.6$, with a metal-poor tail which is thicker than the metal-rich tail, possibly because of a lower-metallicity stellar component at $\textrm{[Fe/H]}\approx -2.2$. This is consistent with the metallicities assigned to an inner halo ($\sim$ 10-15 kpc from the Sun) and an outer halo (beyond $\sim$ 15 kpc from the Sun), respectively \citep{carolloetal2007,carolloetal2010,beersetal2012}. Fig. \ref{fig:comparison_retrievals_others} shows comparisons of our estimates of [Fe/H] based on SDSS spectra, against the SSPP pipeline and values from other recent studies. In the case of the SSPP, the persistent offset may be in part due to the fact that SSPP metallicities depend more heavily on the Ca II K line as the metallicity decreases; and as the metallicity increases more techniques can be applied the find the abundances.

\subsection{Effective Temperature}

We also considered estimates of T$_{\rm eff}$ by using the net Balmer-line width. This relation is highly linear, and a simple fit leads to the relation:

\begin{equation}
    T_{\textrm{eff}} = A\,W(H) + B,
\end{equation}

\begin{figure}
\centering
\includegraphics[width=1.0\linewidth]{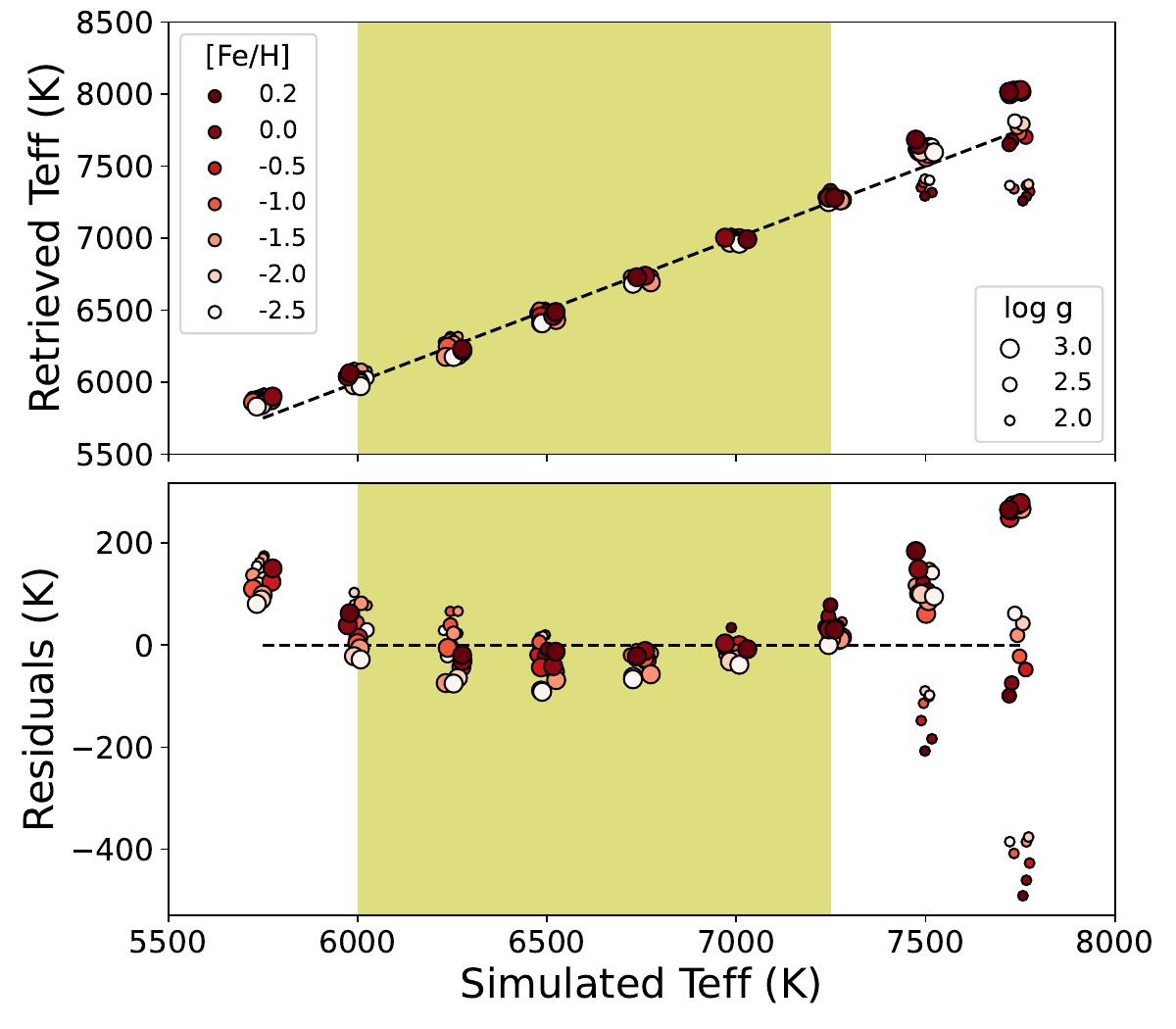}
\caption{Estimated T$_{\rm eff}$ from synthetic spectra, generated at intervals of 250\,K. The yellow band is the approximate boundary of the instability strip. The dashed line is the one-to-one line; some scatter has been introduced for the data points in the x direction to minimize overlap.}
\label{fig:teff_retrieval}
\end{figure}

\begin{figure*}
\centering
\includegraphics[width=1.0\linewidth, trim={0cm 0.6cm 0cm 0cm}, clip=True]{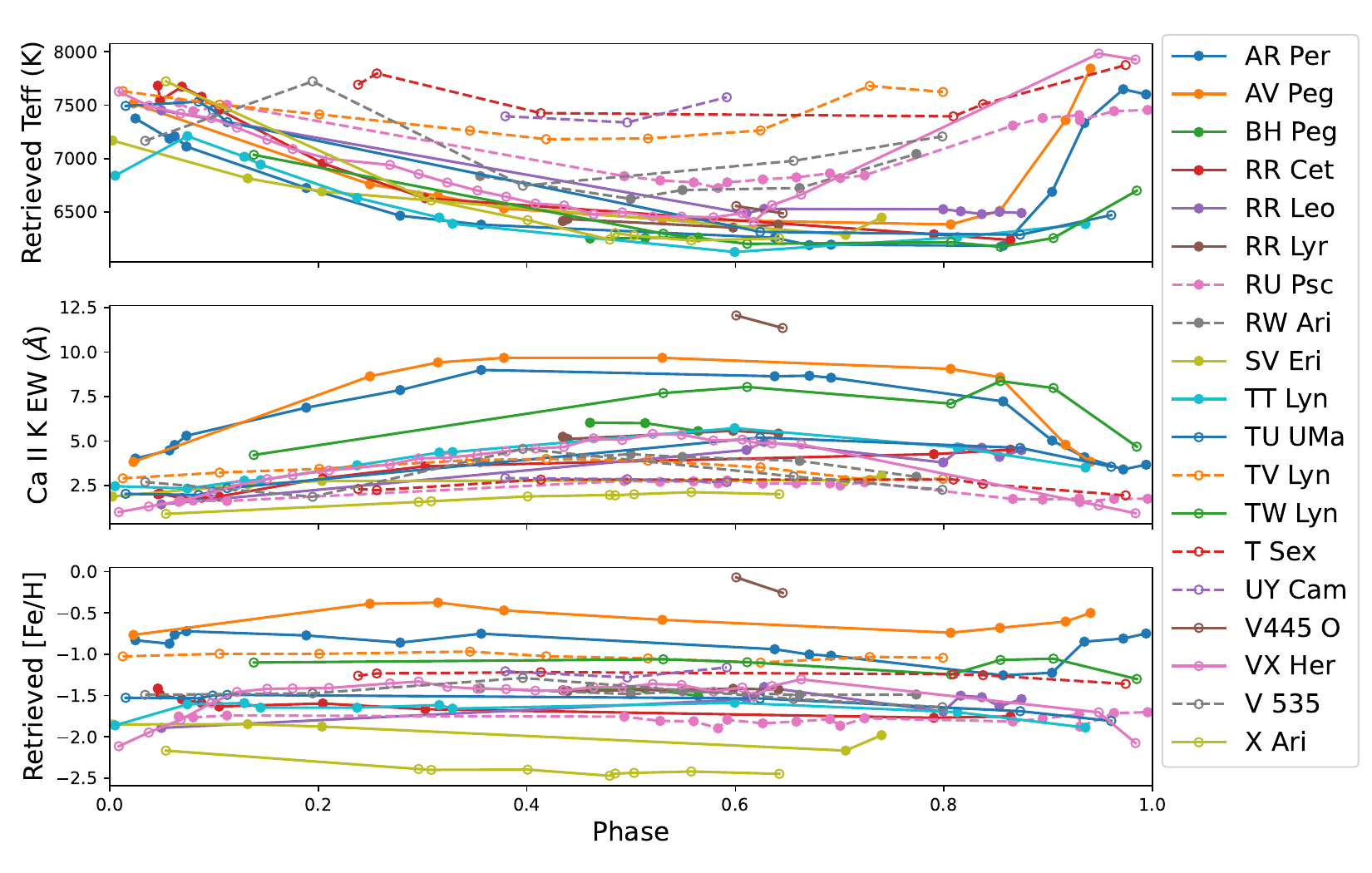}
\caption{Observables of our program stars, as returned by \texttt{rrlfe}. Solid lines are RRab stars, and dashed lines are RRc stars. Qualitatively, the retrieved T$_{\rm eff}$, which is directly proportional to the Balmer EWs, resembles RR Lyrae photometric light curves.
}
\label{fig:temp_mcd}
\end{figure*}

\noindent
where $A=192.8\pm1.6$ K/{\AA}  and $B=5434\pm11$\,K. We apply this to synthetic spectra in Fig.\ \ref{fig:teff_retrieval}. The estimated temperatures are sufficiently accurate that ``temperature curves'' can be obtained, as shown in Fig.\ \ref{fig:temp_mcd}.

\section{Discussion}
\label{sec:secDiscussion}

The calibration we present in this work, and the pipeline used to generate it, is available as a software package (Sec. \ref{sec:data_availability}). It can be applied to SDSS survey spectroscopy, and to refine photometric calibrations such as those that will make use of the LSST dataset. It would also be of interest to find where consistency among these different methods fails.

One advantage of our methodology is that the low-resolution input spectra can be restricted to a range of $\sim$ 3910 -- 4950\,{\AA}, or a bandwidth of $\Delta \lambda\approx 0.1 \mu$m. Our methodology is also not limited to RR Lyrae stars. It can also be used to generate and apply calibrations to main-sequence non-pulsating stars, until the Balmer lines fade in strength at later types. This is not only relevant for studies of Milky Way structure, but also investigations further afield, such as correlations between exoplanet demographics and the metallicities of their host stars. The latter provide clues to the physics of the planet-formation process \citep{kolecki2021searching,boley2021searching,ghezzi2021spectroscopic}, and gleaning metallicities from low-resolution host-star spectra is particularly important now that the sample sizes of known exoplanets has grown far beyond where it is feasible to follow them all up with high-resolution spectroscopy with large telescopes.

There are a number of possible avenues to improve our calibration in the future.  One would be to include more synthetic spectra in the metal-poor ([Fe/H] $< -2.5$) and metal-rich ([Fe/H] $>-1.0$) regimes. This will provide a better lever arm for making refinements based on comparisons with the SSPP pipeline. As a secondary priority, it would be useful to obtain empirical, low-resolution spectra that cover more of the pulsation phase of a given sample of stars. In this work we found consistent metallicities across almost all phases tested, but with more phase coverage the phase region where this calibration breaks down could be constrained further.

To expand the sample of input stars with high-resolution spectroscopy for mapping literature [Fe/H] values, one may consider stars accessible from the Southern Hemisphere, so as to overlap with the 11 field RRab stars of \citet{foretal2011} (see also \citealt{chadidetal2013}). Their stars have unmatched phase coverage with high-resolution ($R\approx 27,000$) spectroscopy, and the calculation by \citet{foretal2011} of stellar parameters like log g would allow us to try to find correlations between them and the metallicities generated by our calibrations.

In this work we have also obtained low-resolution spectra of RRc stars and applied our calibration thereto, but a better comparison of metallicity estimates is needed for more RRc stars with a set of [Fe/H] values from high-resolution spectroscopy. A larger basis set of RRc stars would also allow this subtype to be folded in to the generation of a calibration, or enable a separate RRc calibration. High-priority RRc star targets for low-resolution spectroscopy include those which already appear in high-resolution studies: U Com, T Sex, DH Peg, YZ Cap, and eight southern RRc stars in the All Sky Automated Survey (ASAS) \citep{govea2014chemical}. Target lists that expand beyond these, and which overlap among multiple studies, will be useful for making a mapping specific to RRc stars analogous to our mapping of RRab stars to those in \citetalias{layden1994}.

\section{Summary}
\label{sec:secSummary}

In this work, we have established a metallicity calibration for low- or mid-resolution spectroscopy of RR Lyrae variables from a spectroscopic basis set including both RRab and RRc sub-types. This includes a final correction based on comparison with estimates based on high-resolution spectroscopy in the literature. We make our ``living'' code available, which can be updated with new data, including for RRc stars.

\section{Data availability}
\label{sec:data_availability}

This software is accessible at \url{https://github.com/mwanakijiji/rrlfe}, where issues can be flagged by users. Documentation is at \url{https://rrlfe.readthedocs.io}. Original data underlying this article are available on Zenodo, at \url{https://doi.org/10.5281/zenodo.8053438}.

\newpage
\acknowledgments

Other software: \texttt{matplotlib} \citep{hunter2007matplotlib}, \texttt{numpy} \citep{walt2011numpy}, \texttt{corner} \citep{corner}, \texttt{astropy} \citep{price2018astropy}, Singularity \citep{kurtzer2017singularity}, \texttt{seaborn} \citep{seaborn}.

Acknowledgements: E.S. acknowledges support from Huffaker Travel Scholarships, and department Chair Sumit Das for providing additional travel funding, under the auspices of the University of Kentucky Physics and Astronomy Department. T.C.B. acknowledges partial support for
this work from grant PHY 14-30152; Physics Frontier Center/JINA Center for the Evolution
of the Elements (JINA-CEE), awarded by the U.S. National Science Foundation. The work of V.M.P. is supported by NOIRLab, which is managed by the Association of Universities for Research in Astronomy (AURA) under a cooperative agreement with the National Science Foundation.
Y.S.L. acknowledges support from the National Research Foundation (NRF) of
Korea grant funded by the Ministry of Science and ICT (NRF-2021R1A2C1008679).
Y.S.L. also gratefully acknowledges partial support for his visit to the University
of Notre Dame from OISE-1927130: The International Research Network for Nuclear Astrophysics (IReNA),
awarded by the US National Science Foundation.  
We also acknowledge Rhodes Hart, who obtained a light curve for one of our program stars from Mt. Kent Observatory in Australia, which is administered by the University of Southern Queensland. 

People who provided invaluable assistance for the observations included Kyle McCarthy and Tim Knauer at the MacAdam Student Observatory, and David Doss and John Kuehne at McDonald Observatory. We also thank the University of Arizona's and CyVerse's computing staff members Blake Joyce and Julian Pistorius for their computational assistance. The logo for \texttt{rrlfe} was created by Anna McElhannon. 

Finally, we acknowledge Jim Nemec, Thomas Gomez, and George Wallerstein for kindly providing their high-resolution spectroscopy to constrain line-of-sight interstellar calcium absorption; Chris Waters, for his development of the Python version of \texttt{Robospect}; and Andrew Odell for pointing out the irregular pulsation of RW Ari. 

We acknowledge with thanks the variable star observations from the AAVSO International Database contributed by observers worldwide and used in this research. In particular, we are indebted to observer Gerard Samolyk for his observations of BH Peg and VX Her. Other observers who contributed to the heavy sampling of the VX Her light curve were Teofilo Arranz, Jerry Bialozynski, Neil Butterworth, Shawn Dvorak, Massimiliano Martignoni, Kenneth Menzies, Tom Polakis, John Ritzel, John Ruthroff, Richard Sabo, Neil Simmons, and Denis St-Gelais.

This paper also makes use of data from KELT (DOI 10.26133/NEA8). Early work on KELT-North was supported by NASA Grant NNG04GO70G. Work on KELT-North was partially supported by NSF CAREER Grant AST-1056524 to S. Gaudi, and from the Vanderbilt Office of the Provost through the Vanderbilt Initiative in Data-intensive Astrophysics.

Funding for the SDSS and SDSS-II has been provided by the Alfred P. Sloan Foundation, the Participating Institutions, the National Science Foundation, the U.S. Department of Energy, the National Aeronautics and Space Administration, the Japanese Monbukagakusho, the Max Planck Society, and the Higher Education Funding Council for England. The SDSS Web Site is http://www.sdss.org/.	

The SDSS is managed by the Astrophysical Research Consortium for the Participating Institutions. The Participating Institutions are the American Museum of Natural History, Astrophysical Institute Potsdam, University of Basel, University of Cambridge, Case Western Reserve University, University of Chicago, Drexel University, Fermilab, the Institute for Advanced Study, the Japan Participation Group, Johns Hopkins University, the Joint Institute for Nuclear Astrophysics, the Kavli Institute for Particle Astrophysics and Cosmology, the Korean Scientist Group, the Chinese Academy of Sciences (LAMOST), Los Alamos National Laboratory, the Max-Planck-Institute for Astronomy (MPIA), the Max-Planck-Institute for Astrophysics (MPA), New Mexico State University, Ohio State University, University of Pittsburgh, University of Portsmouth, Princeton University, the United States Naval Observatory, and the University of Washington.

This paper includes data collected by the TESS mission, which are publicly available from the Mikulski Archive for Space Telescopes (MAST). Funding for the TESS mission is provided by NASA’s Science Mission directorate. This research has also made use of the SIMBAD database, operated at CDS, Strasbourg, France \citep{wenger2000simbad}.

This research was supported in part by the Notre Dame Center for Research Computing's computing clusters. Some computations were also performed on CyVerse Atmosphere cyberinfrastructure, which is supported by the National Science Foundation under Award Numbers DBI-0735191 and DBI-1265383, URL: \url{www.cyverse.org}. Jeff Chilcote also kindly allowed the use of his \texttt{planetfinder} cluster.

Author contributions: E.S. led the acquisition and analysis of the spectroscopy and photometry from McDonald Observatory and the MacAdam Student Observatory, the development of the software pipeline, and the preparation of this manuscript. R.W. supervised the effort of E.S., generated synthetic spectra, and contributed to the pipeline. N.D.L. calculated spectral phases with the KELT data and contributed to the pipeline. S.L. obtained SDSS spectroscopy. T.C.B. and V.M.P. ran the n-SSPP on spectra, and inspected the results to ensure quality. J.K. obtained photometry from Moore Observatory. Y.S.L. contributed to updates to the SSPP. J.P. provided and supported KELT survey photometry. K.C. contributed to the pipeline. 

\section{ORCID iDs}

\noindent
Eckhart Spalding {\includegraphics[scale=0.1]{images/orcid_64x64.png}} \url{https://orcid.org/0000-0003-3819-0076} \\
Ronald Wilhelm {\includegraphics[scale=0.1]{images/orcid_64x64.png}} \url{https://orcid.org/0000-0002-4792-7722} \\
Nathan De Lee {\includegraphics[scale=0.1]{images/orcid_64x64.png}} \url{https://orcid.org/0000-0002-3657-0705} \\
Stacy Long {\includegraphics[scale=0.1]{images/orcid_64x64.png}} 
\url{https://orcid.org/0000-0002-0726-424X} \\
Timothy C. Beers {\includegraphics[scale=0.1]{images/orcid_64x64.png}} 
\url{https://orcid.org/0000-0003-4573-6233} \\
Vinicius M. Placco {\includegraphics[scale=0.1]{images/orcid_64x64.png}} 
\url{https://orcid.org/0000-0003-4479-1265} \\
John Kielkopf {\includegraphics[scale=0.1]{images/orcid_64x64.png}} \url{https://orcid.org/0000-0003-0497-2651} \\
Young Sun Lee {\includegraphics[scale=0.1]{images/orcid_64x64.png}} \url{https://orcid.org/0000-0001-5297-4518} \\
Joshua Pepper {\includegraphics[scale=0.1]{images/orcid_64x64.png}} \url{https://orcid.org/0000-0002-3827-8417} \\
Kenneth Carrell {\includegraphics[scale=0.1]{images/orcid_64x64.png}} \url{https://orcid.org/0000-0002-6307-992X} \\

\facilities{McDonald, Struve, Sloan, AAVSO, KELT, TESS}

\appendix
\setcounter{table}{0}
\renewcommand{\thetable}{A\arabic{table}}

\section{Program star metallicities}
\label{sec:prog_star_feh}

Table \ref{table:sources} lists the literature sources of [Fe/H] based on high-resolution spectroscopy for our program stars, upon which we make a basis set to map our own estimated values. In addition, we list \citetalias{layden1994} upon which a consistent mapping between literature sources is based. Table \ref{table:prog_star_fehs} lists both the mapped and estimated values for each star.

\begin{longrotatetable}
\begin{deluxetable*}{l |l |p{1.5cm} |p{2cm}|p{12cm}}
\tablecaption{Literature [Fe/H] Sources \label{table:sources}}
\tablecolumns{5}
\tabletypesize{\footnotesize}
\tablehead{
Reference     & Abbrev.            & $R$/1000 & Lines   & Remarks   
}
\startdata
\citet{layden1994} & L94 & Dispersion of 1-9\,{\AA} per channel & Ca II K and Balmer lines used to calibrate to a [Fe/H] scale &  This source serves as our basis set for RRab stars. Abundances were determined by using the method of \citet{freeman1975chemical}, which measures EWs. \\
\hline
\citet{clementinietal1995} & C95 & $\sim38$ & FeI, FeII &  Used spectra near light curve minimum, and found $\textrm{[Fe/H]}$ values based on the average of $\textrm{log}(\epsilon)$ from FeI and FeII lines.  \\
\hline
\citet{lambertetal1996} & L96 & $\sim23$ & FeII and photometric models &  LTE is assumed. Values of $\textrm{log}(\epsilon)$ in their Table 3 converted to [Fe/H] using solar value of $7.51\pm0.01$ from \citet{anstee1997determination}. See \citetalias{lambertetal1996} Sec.\  3.2. \\
\hline
\citet{fernleyetal1997} & F97 & $\sim60$ & FeII &  Used spectra near light curve minimum. Abundances are from electronic appendix on CDS. See Sec.\  4.1 in \citetalias{fernleyetal1997}. (Note this work supercedes \citet{fernleyetal1996}.) \\
\hline
\citet{solanoetal1997} & S97 & $\sim20$ & FeI & Multiple [Fe/H] values from different phases. Abundances are posted in electronic appendix on CDS.  (See \citetalias{solanoetal1997} Sec.\  4.) \\
\hline
\citet{wallerstein2010composition} & W10 & $\sim30$ & FeII & See \citetalias{wallerstein2010composition} Table 2 and Sec.\  2. \\
\hline
\citet{liu2013abundances} & L13 & $\sim60$ & FeI, FeII & Some stars observed at multiple phases. We take the average of abundances in those cases. See \citetalias{liu2013abundances} Tables 3, 4, 5.  \\
\hline
\citet{nemecetal2013} & N13 & $\sim65$ and $\sim36$ & FeI, FeII & Abundances are the weighted average of abundances from Fe I and Fe II lines. See \citetalias{nemecetal2013} Table 7. \\
\hline
\citet{pancino2015chemical} & P15 & $\geq30$ & FeI & Spectra are taken at different phases. See \citetalias{pancino2015chemical}  Table 7.\\
\hline
\citet{chadid2017spectroscopic} & C17 & $\sim27$ & FeI, FeII & Used spectra near light curve minimum. We use abundances which are the average of those of the Fe I and Fe II lines. See \citetalias{chadid2017spectroscopic} Table 3.\\
\hline
\citet{sneden2017rrc} & S17 & $\sim27$ & FeI, FeII & We use a single star from \citetalias{sneden2017rrc}, T Sex, for which they obtained spectra at phases 0.14 to 0.62. See \citetalias{sneden2017rrc} Tables 1, 4 and 6. \\
\hline
\citet{crestani2021use} & C21 & $\sim35$ & FeI, FeII & Obtained spectra across all pulsation phases, with EWs measured manually. Some spectra are stacked at similar pulsation phase to increase S/N. See \citetalias{C21} Sec.\  3.1. \\ 
\enddata
\end{deluxetable*}
\end{longrotatetable}

\begin{longrotatetable}
\begin{deluxetable*}{l l p{2cm} r r | p{3cm} | p{3cm} |}
\tablecaption{Mapped Metallicities \label{table:prog_star_fehs}}
\tablecolumns{6}
\tabletypesize{\footnotesize}
\tablehead{
Star     & Subtype            & Literature [Fe/H] & Error (dex)   & Ref  & [Fe/H], mapped literature and L94 basis for RRab stars & [Fe/H], our calibration$^{a}$}
\startdata
RW Ari 		& c & 	-1.48 	& 	$\pm$0.25	& 	\citetalias{kemper1982}	                    & -       & -1.47$\pm$0.01$\pm$0.03	\\
\hline
X Ari 		& ab &	-2.50 	& 	$\pm$0.09	& \citetalias{clementinietal1995}               & -2.56$\pm$0.12  & -2.40$\pm$0.01$\pm$0.09	\\
`` `` 		& `` &	-2.48 	& 	$\pm$0.13	&  	\citetalias{lambertetal1996}                &                             & 	\\
`` `` 		& `` &	-2.40 	& 	$\pm$0.20	& 	 	\citetalias{layden1994}                 & 	                            &	\\
`` `` 		& `` &	-2.19 	& 	$\pm$0.17	&  	\citetalias{pancino2015chemical}            & &	\\
`` `` 		& `` &	-2.60 	& 	$\pm$0.11	&  	\citetalias{chadid2017spectroscopic}        & & 	\\
`` `` 		& `` &	-2.68 	& 	$\pm$0.18	& 	 	\citetalias{wallerstein2010composition}	& &	\\
`` `` 		& `` &	-2.74 	& 	$\pm$0.09	& 	\citetalias{nemecetal2013} 	                & & 	\\
\hline
UY Cam 		& 	c 	& 	-1.51	& 	$\pm$0.13	& 	\citetalias{fernleyetal1997} 	        & - & -1.22$\pm$0.01$\pm$0.06 	\\
`` `` 		& 	`` 	& 	-1.06	& 	$\pm$0.25	& 	\citetalias{kemper1982} 	            & &  	\\
\hline
RR Cet 		& 	ab 	& 	-1.38	& 	$\pm$0.09	& 	\citetalias{clementinietal1995} 	    & -1.64$\pm$0.14 & -1.61$\pm$0.01$\pm$0.12  	\\
`` ``  		& 	`` 	& 	-1.49	& 	$\pm$0.13	& 	\citetalias{fernleyetal1997}            & & 	 	\\
`` ``  		& 	`` 	& 	-1.52	& 	$\pm$0.20	& 	\citetalias{layden1994}                 & & 	 	\\
`` ``  		& 	`` 	& 	-1.31	& 	$\pm$0.18	& 	\citetalias{solanoetal1997}	            & & 	\\
`` ``  		& 	`` 	& 	-1.57	& 	$\pm$0.11	& 	\citetalias{chadid2017spectroscopic}    & & 	 	\\
`` `` 		& 	`` 	& 	-1.61	& 	$\pm$0.18 & 	\citetalias{wallerstein2010composition}	& &  	\\
`` ``  		& 	`` 	& 	-1.18	& 	$\pm$0.09	& 	\citetalias{nemecetal2013}              & & 	 	\\
`` ``  		& 	`` 	& 	-1.35	& 	$\pm$0.13	& 	\citetalias{liu2013abundances}          & & 	\\
\hline
SV Eri  		& 	ab 	& 	-2.11	& 	$\pm$0.13	& 	\citetalias{lambertetal1996} 	    & -2.18$\pm$0.10 & -1.94$\pm$0.01$\pm$0.14	\\
`` ``  		& 	`` 	& 	-2.04	& 	$\pm$0.20	& 	\citetalias{layden1994}	                & & 	\\
`` ``  		& 	`` 	& 	-1.94	& 	$\pm$0.18	& 	\citetalias{wallerstein2010composition}	& & 	\\
\hline
VX Her 		& ab &	-1.58 	& 	$\pm$0.09	& 	\citetalias{clementinietal1995}             & -- & -1.51$\pm$0.01$\pm$0.23 	 	\\
`` ``		& `` &	-1.56 	& 	$\pm$0.17	& 	\citetalias{pancino2015chemical}	        & &  	\\
`` ``		& `` &	-1.48 	& 	$\pm$0.18	& 	\citetalias{wallerstein2010composition}     & & 	 	\\
`` `` 		& `` &	-1.23 	& 	$\pm$0.12	& 	\citetalias{nemecetal2013}                  & & 	 	\\
`` `` 		& `` &	-1.33 	& 	$\pm$0.13	& 	\citetalias{liu2013abundances}              & & 	 	\\
\hline
RR Leo 		& ab 	& 	-1.40	& 	$\pm$0.13	& 	\citetalias{lambertetal1996}            & -1.69$\pm$0.20 & -1.59$\pm$0.01$\pm$0.15	 	\\
`` `` 		& 	``  & 	-1.57	& 	$\pm$0.20	& 	\citetalias{layden1994}                 & & 	 	\\
`` `` 		& 	``  & 	-1.39	& 	$\pm$0.18	& 	\citetalias{wallerstein2010composition} & & 	 	\\
\hline
TT Lyn 		& ab &	-1.53 	& 	$\pm$0.13	& 		\citetalias{fernleyetal1997}            & -- & -1.68$\pm$0.01$\pm$0.11 	\\
`` `` 		& 	`` &	-1.55 	& 	$\pm$0.13	&  	\citetalias{lambertetal1996}            & & 	 	\\
`` `` 		& 	`` &	-1.34 	& 	$\pm$0.18	& 	\citetalias{solanoetal1997}	            &  & 	 	\\
`` `` 		& 	`` &	-1.41 	& 	$\pm$0.18	& 	\citetalias{wallerstein2010composition}	&  & 	 	\\
\hline
TV Lyn 		& c	 	& 	-0.99	& 	$\pm$0.25	& 	\citetalias{kemper1982}                 & - & -1.03$\pm$0.01$\pm$0.04	 	\\
\hline
TW Lyn 		& 	ab 	& 	-0.09	& 	$\pm$0.13	& 	\citetalias{fernleyetal1997}            & -1.34$\pm$0.26 & -1.13$\pm$0.01$\pm$0.10	 	\\
`` ``  		& 	`` 	& 	-1.23	& 	$\pm$0.20	& 	\citetalias{layden1994}                 & & 	 	\\
\hline
RR Lyr 		& ab &	-1.39 	& 	$\pm$0.09	& 	\citetalias{clementinietal1995} 	        &  -1.49$\pm$0.10 & -1.43$\pm$0.01$\pm$0.01	 	\\
`` `` 		& `` 	&	-1.37 	& 	$\pm$0.20	& 	\citetalias{layden1994}	                &  & 	 	\\
`` `` 		& `` 	&	-1.44 	& 	$\pm$0.18 & 	\citetalias{wallerstein2010composition}	&  & 	 	\\
`` `` 		& `` 	&	-1.27 	& 	$\pm$0.12	& 	\citetalias{nemecetal2013} 	            &  & 	 	\\
`` `` 		& `` 	&	-1.39 	& 	$\pm$0.13	& 	\citetalias{liu2013abundances} 	        &  & 	 	\\
\hline
V535 Mon 		& 	c 	& 	-1.44	& 	$\pm$0.25	& 	\citetalias{kemper1982}             & - & -1.49$\pm$0.01$\pm$0.13	 	\\
\hline
V445 Oph 		& 	ab 	& 	+0.17	& 	$\pm$0.12	& 	\citetalias{clementinietal1995}     & -0.30$\pm$0.41 & -0.16$\pm$0.02$\pm$0.13	 	\\
`` `` 		& 	`` 	& 	-0.23	& 	$\pm$0.20	& 	\citetalias{layden1994}                 & & 	 	\\
`` ``  		& 	`` 	& 	-0.12	& 	$\pm$0.11	& 	\citetalias{chadid2017spectroscopic}    & & 	 	\\
`` `` 		& 	`` 	& 	+0.24	& 	$\pm$0.18	& 	\citetalias{wallerstein2010composition} & & 	 	\\
`` `` 		& 	`` 	& 	+0.13	& 	$\pm$0.10	& 	\citetalias{nemecetal2013}              & & 	 	\\
`` `` 		& 	`` 	& 	+0.14	& 	$\pm$0.13	& 	\citetalias{liu2013abundances}          & & 	 	\\
\hline
AV Peg 		& 	ab 	& 	-0.22	& 	$\pm$0.13	& 	\citetalias{lambertetal1996}            & -0.21$\pm$0.23 & -0.57$\pm$0.01$\pm$0.14 	\\
`` ``		& 	`` 	& 	-0.14	& 	$\pm$0.20	& 	\citetalias{layden1994}                 & & 	 	\\
`` ``		& 	`` 	& 	-0.17	& 	$\pm$0.11	& 	\citetalias{chadid2017spectroscopic}    & & 	 	\\
\hline
BH Peg 		& 	ab 	& 	-1.38	& 	$\pm$0.20	& 	\citetalias{layden1994}                 & -1.50$\pm$0.22 & -1.45$\pm$0.01$\pm$0.05	 	\\
`` `` 		& 	`` 	& 	-1.17	& 	$\pm$0.18	& 	\citetalias{wallerstein2010composition} & & 	 	\\
\hline
AR Per 		& 	ab 	& 	-0.29	& 	$\pm$0.13	& 	\citetalias{fernleyetal1997}            & -0.51$\pm$0.14 & -0.90$\pm$0.01$\pm$0.17	 	\\
`` `` 		& 	`` 	& 	-0.24	& 	$\pm$0.13	& 	\citetalias{lambertetal1996}            & & 	 	\\
`` `` 		& 	`` 	& 	-0.43	& 	$\pm$0.20	& 	\citetalias{layden1994}                 & & 	 	\\
`` `` 		& 	`` 	& 	-0.23	& 	$\pm$0.18	& 	\citetalias{solanoetal1997}             & & 	 	\\
`` `` 		& 	`` 	& 	-0.32	& 	$\pm$0.18 & 	\citetalias{wallerstein2010composition} & & 	 	\\
\hline
RU Psc 		& 	c 	& 	-1.65	& 	$\pm$0.25	& 	\citetalias{kemper1982}                 & - & -1.79$\pm$0.01$\pm$0.05	 	\\
`` `` 		& 	`` 	& 	-2.04	& 	$\pm$0.18	& 	\citetalias{wallerstein2010composition} & & 	 	\\
\hline
T Sex 		& 	c 	& 	-1.37	& 	$\pm$0.13	& 	\citetalias{fernleyetal1997}            & - & -1.26$\pm$0.01$\pm$0.05	 	\\
`` `` 		& 	- 	& 	-1.18	& 	$\pm$0.25	&  \citetalias{kemper1982}                  & & 	 	\\
`` `` 		& 	- 	& 	-1.56	& 	$\pm$0.13	& 	\citetalias{lambertetal1996}            & & 	 	\\
`` `` 		& 	- 	& 	-1.27	& 	$\pm$0.18	& 	\citetalias{solanoetal1997}             & & 	 	\\
\hline
TU UMa 		& 	ab 	& 	-1.57	& 	$\pm$0.13	& 	\citetalias{fernleyetal1997}            & -1.56$\pm$0.17 & -1.58$\pm$0.01$\pm$0.12 	\\
`` `` 		& 	`` 	& 	-1.56	& 	$\pm$0.13	& 	\citetalias{lambertetal1996}            & & 	 	\\
`` ``		& 	`` 	& 	-1.44	& 	$\pm$0.20	& 	\citetalias{layden1994}                 & & 	 	\\
`` ``		& 	`` 	& 	-1.31	& 	$\pm$0.05	& 	\citetalias{pancino2015chemical}        & & 	 	\\
`` `` 		& 	`` 	& 	-1.46	& 	$\pm$0.18	& 	\citetalias{wallerstein2010composition} & & 	 	\\
\enddata
\tablecomments{
$^a$~The first error term is the average error in [Fe/H] across all spectra for that star. The second is the scatter in [Fe/H] between spectra of the same star. 
}
\end{deluxetable*}
\end{longrotatetable}

\begin{longrotatetable}
\begin{deluxetable*}{l l p{2cm} r r | p{3cm} | p{3cm} |}
\tablecaption{Mapped Metallicities \label{table:prog_star_fehs}}
\tablecolumns{6}
\tabletypesize{\footnotesize}
\tablehead{
Star     & Subtype            & Literature [Fe/H] & Error (dex)   & Ref  & [Fe/H], mapped literature and L94 basis for RRab stars & [Fe/H], our calibration$^{a}$}
\startdata
RW Ari 		& c & 	-1.48 	& 	$\pm$0.25	& 	\citetalias{kemper1982}	                    & -       & -1.47$\pm$0.01$\pm$0.03	\\
\hline
X Ari 		& ab &	-2.50 	& 	$\pm$0.09	& \citetalias{clementinietal1995}               & -2.56$\pm$0.12  & -2.40$\pm$0.01$\pm$0.09	\\
`` `` 		& `` &	-2.48 	& 	$\pm$0.13	&  	\citetalias{lambertetal1996}                &                             & 	\\
`` `` 		& `` &	-2.40 	& 	$\pm$0.20	& 	 	\citetalias{layden1994}                 & 	                            &	\\
`` `` 		& `` &	-2.19 	& 	$\pm$0.17	&  	\citetalias{pancino2015chemical}            & &	\\
`` `` 		& `` &	-2.60 	& 	$\pm$0.11	&  	\citetalias{chadid2017spectroscopic}        & & 	\\
`` `` 		& `` &	-2.68 	& 	$\pm$0.18	& 	 	\citetalias{wallerstein2010composition}	& &	\\
`` `` 		& `` &	-2.74 	& 	$\pm$0.09	& 	\citetalias{nemecetal2013} 	                & & 	\\
\hline
UY Cam 		& 	c 	& 	-1.51	& 	$\pm$0.13	& 	\citetalias{fernleyetal1997} 	        & - & -1.22$\pm$0.01$\pm$0.06 	\\
`` `` 		& 	`` 	& 	-1.06	& 	$\pm$0.25	& 	\citetalias{kemper1982} 	            & &  	\\
\hline
RR Cet 		& 	ab 	& 	-1.38	& 	$\pm$0.09	& 	\citetalias{clementinietal1995} 	    & -1.64$\pm$0.14 & -1.61$\pm$0.01$\pm$0.12  	\\
`` ``  		& 	`` 	& 	-1.49	& 	$\pm$0.13	& 	\citetalias{fernleyetal1997}            & & 	 	\\
`` ``  		& 	`` 	& 	-1.52	& 	$\pm$0.20	& 	\citetalias{layden1994}                 & & 	 	\\
`` ``  		& 	`` 	& 	-1.31	& 	$\pm$0.18	& 	\citetalias{solanoetal1997}	            & & 	\\
`` ``  		& 	`` 	& 	-1.57	& 	$\pm$0.11	& 	\citetalias{chadid2017spectroscopic}    & & 	 	\\
`` `` 		& 	`` 	& 	-1.61	& 	$\pm$0.18 & 	\citetalias{wallerstein2010composition}	& &  	\\
`` ``  		& 	`` 	& 	-1.18	& 	$\pm$0.09	& 	\citetalias{nemecetal2013}              & & 	 	\\
`` ``  		& 	`` 	& 	-1.35	& 	$\pm$0.13	& 	\citetalias{liu2013abundances}          & & 	\\
\hline
SV Eri  		& 	ab 	& 	-2.11	& 	$\pm$0.13	& 	\citetalias{lambertetal1996} 	    & -2.18$\pm$0.10 & -1.94$\pm$0.01$\pm$0.14	\\
`` ``  		& 	`` 	& 	-2.04	& 	$\pm$0.20	& 	\citetalias{layden1994}	                & & 	\\
`` ``  		& 	`` 	& 	-1.94	& 	$\pm$0.18	& 	\citetalias{wallerstein2010composition}	& & 	\\
\hline
VX Her 		& ab &	-1.58 	& 	$\pm$0.09	& 	\citetalias{clementinietal1995}             & -- & -1.51$\pm$0.01$\pm$0.23 	 	\\
`` ``		& `` &	-1.56 	& 	$\pm$0.17	& 	\citetalias{pancino2015chemical}	        & &  	\\
`` ``		& `` &	-1.48 	& 	$\pm$0.18	& 	\citetalias{wallerstein2010composition}     & & 	 	\\
`` `` 		& `` &	-1.23 	& 	$\pm$0.12	& 	\citetalias{nemecetal2013}                  & & 	 	\\
`` `` 		& `` &	-1.33 	& 	$\pm$0.13	& 	\citetalias{liu2013abundances}              & & 	 	\\
\hline
RR Leo 		& ab 	& 	-1.40	& 	$\pm$0.13	& 	\citetalias{lambertetal1996}            & -1.69$\pm$0.20 & -1.59$\pm$0.01$\pm$0.15	 	\\
`` `` 		& 	``  & 	-1.57	& 	$\pm$0.20	& 	\citetalias{layden1994}                 & & 	 	\\
`` `` 		& 	``  & 	-1.39	& 	$\pm$0.18	& 	\citetalias{wallerstein2010composition} & & 	 	\\
\hline
TT Lyn 		& ab &	-1.53 	& 	$\pm$0.13	& 		\citetalias{fernleyetal1997}            & -- & -1.68$\pm$0.01$\pm$0.11 	\\
`` `` 		& 	`` &	-1.55 	& 	$\pm$0.13	&  	\citetalias{lambertetal1996}            & & 	 	\\
`` `` 		& 	`` &	-1.34 	& 	$\pm$0.18	& 	\citetalias{solanoetal1997}	            &  & 	 	\\
`` `` 		& 	`` &	-1.41 	& 	$\pm$0.18	& 	\citetalias{wallerstein2010composition}	&  & 	 	\\
\hline
TV Lyn 		& c	 	& 	-0.99	& 	$\pm$0.25	& 	\citetalias{kemper1982}                 & - & -1.03$\pm$0.01$\pm$0.04	 	\\
\hline
TW Lyn 		& 	ab 	& 	-0.09	& 	$\pm$0.13	& 	\citetalias{fernleyetal1997}            & -1.34$\pm$0.26 & -1.13$\pm$0.01$\pm$0.10	 	\\
`` ``  		& 	`` 	& 	-1.23	& 	$\pm$0.20	& 	\citetalias{layden1994}                 & & 	 	\\
\hline
RR Lyr 		& ab &	-1.39 	& 	$\pm$0.09	& 	\citetalias{clementinietal1995} 	        &  -1.49$\pm$0.10 & -1.43$\pm$0.01$\pm$0.01	 	\\
`` `` 		& `` 	&	-1.37 	& 	$\pm$0.20	& 	\citetalias{layden1994}	                &  & 	 	\\
`` `` 		& `` 	&	-1.44 	& 	$\pm$0.18 & 	\citetalias{wallerstein2010composition}	&  & 	 	\\
`` `` 		& `` 	&	-1.27 	& 	$\pm$0.12	& 	\citetalias{nemecetal2013} 	            &  & 	 	\\
`` `` 		& `` 	&	-1.39 	& 	$\pm$0.13	& 	\citetalias{liu2013abundances} 	        &  & 	 	\\
\hline
V535 Mon 		& 	c 	& 	-1.44	& 	$\pm$0.25	& 	\citetalias{kemper1982}             & - & -1.49$\pm$0.01$\pm$0.13	 	\\
\hline
V445 Oph 		& 	ab 	& 	+0.17	& 	$\pm$0.12	& 	\citetalias{clementinietal1995}     & -0.30$\pm$0.41 & -0.16$\pm$0.02$\pm$0.13	 	\\
`` `` 		& 	`` 	& 	-0.23	& 	$\pm$0.20	& 	\citetalias{layden1994}                 & & 	 	\\
`` ``  		& 	`` 	& 	-0.12	& 	$\pm$0.11	& 	\citetalias{chadid2017spectroscopic}    & & 	 	\\
`` `` 		& 	`` 	& 	+0.24	& 	$\pm$0.18	& 	\citetalias{wallerstein2010composition} & & 	 	\\
`` `` 		& 	`` 	& 	+0.13	& 	$\pm$0.10	& 	\citetalias{nemecetal2013}              & & 	 	\\
`` `` 		& 	`` 	& 	+0.14	& 	$\pm$0.13	& 	\citetalias{liu2013abundances}          & & 	 	\\
\hline
AV Peg 		& 	ab 	& 	-0.22	& 	$\pm$0.13	& 	\citetalias{lambertetal1996}            & -0.21$\pm$0.23 & -0.57$\pm$0.01$\pm$0.14 	\\
`` ``		& 	`` 	& 	-0.14	& 	$\pm$0.20	& 	\citetalias{layden1994}                 & & 	 	\\
`` ``		& 	`` 	& 	-0.17	& 	$\pm$0.11	& 	\citetalias{chadid2017spectroscopic}    & & 	 	\\
\hline
BH Peg 		& 	ab 	& 	-1.38	& 	$\pm$0.20	& 	\citetalias{layden1994}                 & -1.50$\pm$0.22 & -1.45$\pm$0.01$\pm$0.05	 	\\
`` `` 		& 	`` 	& 	-1.17	& 	$\pm$0.18	& 	\citetalias{wallerstein2010composition} & & 	 	\\
\hline
AR Per 		& 	ab 	& 	-0.29	& 	$\pm$0.13	& 	\citetalias{fernleyetal1997}            & -0.51$\pm$0.14 & -0.90$\pm$0.01$\pm$0.17	 	\\
`` `` 		& 	`` 	& 	-0.24	& 	$\pm$0.13	& 	\citetalias{lambertetal1996}            & & 	 	\\
`` `` 		& 	`` 	& 	-0.43	& 	$\pm$0.20	& 	\citetalias{layden1994}                 & & 	 	\\
`` `` 		& 	`` 	& 	-0.23	& 	$\pm$0.18	& 	\citetalias{solanoetal1997}             & & 	 	\\
`` `` 		& 	`` 	& 	-0.32	& 	$\pm$0.18 & 	\citetalias{wallerstein2010composition} & & 	 	\\
\hline
RU Psc 		& 	c 	& 	-1.65	& 	$\pm$0.25	& 	\citetalias{kemper1982}                 & - & -1.79$\pm$0.01$\pm$0.05	 	\\
`` `` 		& 	`` 	& 	-2.04	& 	$\pm$0.18	& 	\citetalias{wallerstein2010composition} & & 	 	\\
\hline
T Sex 		& 	c 	& 	-1.37	& 	$\pm$0.13	& 	\citetalias{fernleyetal1997}            & - & -1.26$\pm$0.01$\pm$0.05	 	\\
`` `` 		& 	- 	& 	-1.18	& 	$\pm$0.25	&  \citetalias{kemper1982}                  & & 	 	\\
`` `` 		& 	- 	& 	-1.56	& 	$\pm$0.13	& 	\citetalias{lambertetal1996}            & & 	 	\\
`` `` 		& 	- 	& 	-1.27	& 	$\pm$0.18	& 	\citetalias{solanoetal1997}             & & 	 	\\
%`` ``		& 	- 	& 	(see email from RW, Aug 6, 2019)	& 	(see doc)	& 	\citetalias{sneden2017rrc} & & 	 	\\ # removed from table, since this is the only data point from Sneden
\hline
TU UMa 		& 	ab 	& 	-1.57	& 	$\pm$0.13	& 	\citetalias{fernleyetal1997}            & -1.56$\pm$0.17 & -1.58$\pm$0.01$\pm$0.12 	\\
`` `` 		& 	`` 	& 	-1.56	& 	$\pm$0.13	& 	\citetalias{lambertetal1996}            & & 	 	\\
`` ``		& 	`` 	& 	-1.44	& 	$\pm$0.20	& 	\citetalias{layden1994}                 & & 	 	\\
`` ``		& 	`` 	& 	-1.31	& 	$\pm$0.05	& 	\citetalias{pancino2015chemical}        & & 	 	\\
`` `` 		& 	`` 	& 	-1.46	& 	$\pm$0.18	& 	\citetalias{wallerstein2010composition} & & 	 	\\
\enddata
\tablecomments{
$^a$~The first error term is the average error in [Fe/H] across all spectra for that star. The second is the scatter in [Fe/H] between spectra of the same star. 
}
\end{deluxetable*}
\end{longrotatetable}

\section{Function fitting}
\label{sec:secFitting}

The mapped metallicities for our program stars (see Sec.\  \ref{subsec:lit_metal} and Table \ref{table:prog_star_fehs}) and the Ca II K and Balmer EWs were fed into the MCMC package \texttt{emcee} to fit a variation of the \citetalias{layden1994} relation:

\begin{equation}
W(K)=a + b\,W(H)+c\,\textrm{[Fe/H]}+d\,W(H)\textrm{[Fe/H]},
\label{eqn:layden}
\end{equation}

We fit an expansion of the relation up to third order. Using the shorthand $F$ for [Fe/H], $H$ for W(H), and $K$ for W(K):

\begin{equation} \label{eqn:expanded_lay}
\begin{split}
K= a&+b\,H+c\,F+d\,HF\\
&+f\,H^{2}+g\,F^{2} +h\,FH^{2}+k\,HF^{2}\\
&+m\,H^{3}+n\,F^{3},\\
\end{split}
\end{equation}

\noindent
with coefficients $\{a,b,c,d\}$ all non-zero, and all possible zero/non-zero permutations of the extra six coefficients $\{f,g,h,k,m,n\}$. The number of permutations is the total of all choices of ``6 pick $k$, for $k$ from zero to 6'':

\begin{equation}
\sum_{i=0}^{6} \begin{pmatrix} 6 \\ k \end{pmatrix} = \sum_{i=0}^{6} \frac{6!}{k!(6-k)!} = 64.
\label{eqn:pick}
\end{equation}

\noindent
These 64 combinations include the null set of extra coefficients, which is equivalent to the L94 relation. The validity of each model was tested with the Bayesian Information Criterion (BIC) \citep{schwarz1978estimating} defined as:

\begin{equation}
\textrm{BIC} \equiv \mathscr{k}\, \textrm{ln}(\mathscr{n}) -2\,\textrm{ln}(\mathscr{L}),
\end{equation}

\noindent
where $\mathscr{k}$ is the number of free parameters, $n$ is the number of data points, and $\mathscr{L}$ is the model likelihood function. We have taken $\mathscr{L} \propto exp(-\frac{1}{2}\chi^{2})$, where $\chi^{2}$ is the sum of the weighted squared residuals between the measured equivalent width of the Ca II K line and that retrieved by using the model 
Eqn.~\ref{eqn:expanded_lay} with the best-fitting  coefficients. The sum is taken over spectra $i$, weighted by the error $\sigma_{K0}$ in the Ca II K line returned by \texttt{Robospect}:

\begin{equation}
\chi^{2} = \sum_{i}\frac{[ K_{0,i}-K_{m,i}(\pmb{x_{i}}; \pmb{\theta_{i}}) ]^{2}}{\sigma_{K0,i}^{2}},
\end{equation}

\noindent
where $K_{0}$ is measured, and $K_{m,i}$ is the model value. This value depends on the Balmer-line EW and metallicity, or $\pmb{x_{i}}=\{H_{i},F_{i}\}$; $\pmb{\theta}$ is the vector of coefficients as defined in Eqn.~\ref{eqn:expanded_lay}.

\begin{figure}
\centering
\includegraphics[width=1.0\linewidth]{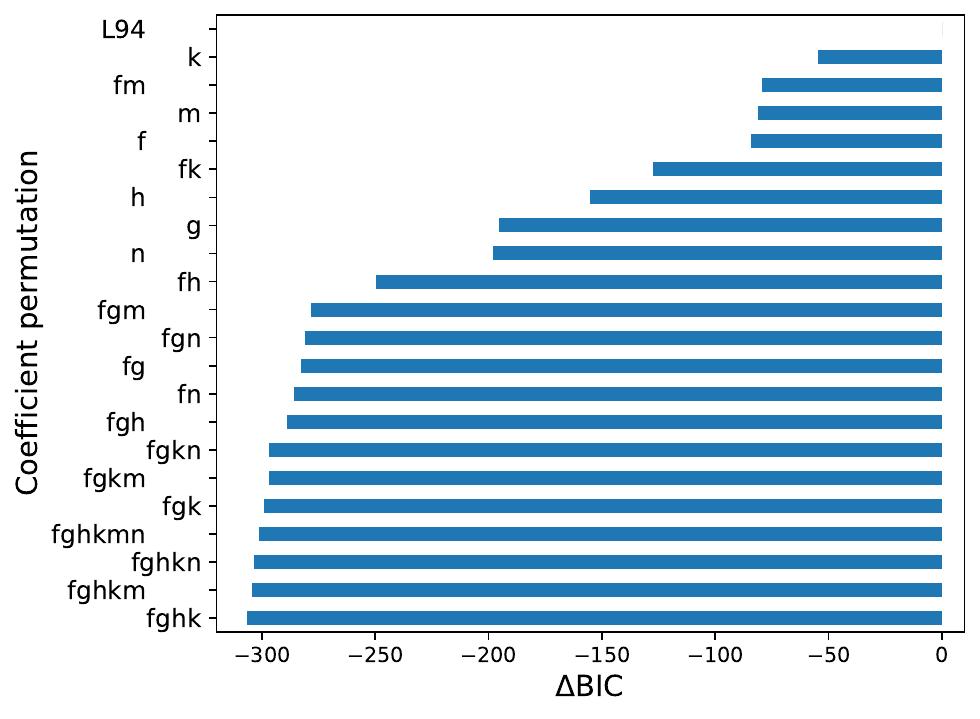}
\caption{Improvements on the \citet{layden1994} model fit (`L94'), based on Eqn.~\ref{eqn:expanded_lay} and different permutations of coefficients $\{f,g,h,k,m,n\}$. (For example, `fgh' is the fit of Eqn.~\ref{eqn:expanded_lay} with $k=m=n=0$.) More negative values of $\Delta$BIC represent an improvement.}
\label{fig:bic_improvement}
\end{figure}

We fit models representing every permutation in Eqn.~\ref{eqn:pick} and calculated the BIC. The difference $\Delta \textrm{BIC}$ between one model and L94 compares the fits while penalizing extra model parameters. Fig.~\ref{fig:bic_improvement} shows all models that converged, and were an improvement on L94. We found the best-fitting  model to be Eqn.~\ref{eqn:expanded_lay} with $m=n=0$ and all other coefficients non-zero.

The solution to Eqn.~\ref{eqn:expanded_lay} with $m=n=0$ for the metallicity $F$ is simply the the quadratic formula:

\begin{equation}
F=\frac{-B\pm \sqrt{B^{2}-4AC}}{2A},
\label{eqn:f_quad}
\end{equation}

\noindent
where 

\begin{equation}
A=g +k\,H, \hspace{0.5cm} B=c +d\,H+hH^{2}, \hspace{0.5cm} C = a +b\,H+\,fH^{2} -K.
\end{equation}

\noindent
Eqn.~\ref{eqn:f_quad} has two solutions, of which the `+' solution in the numerator is the physical one.

\bibliographystyle{apj}
\bibliography{rrl_refs}

\end{document}